\documentclass[12pt, reqno, centertags, titlepage, oneside]{amsart}
\usepackage{amsmath}
\usepackage{amsfonts}
\usepackage{amssymb}
\usepackage{enumerate}
\usepackage{array}
\usepackage{graphicx}

\begin{document}

\title{A Gibbsian Approach to Potential Game Theory}
\author{Michael Campbell\\ \\
	  version 8.1 Feb 2005} 
\address{Innovative Research Concepts, Anaheim, CA, USA}
\email{mcampbel123@yahoo.com}
\begin{abstract}
In games for which there exists a potential, the deviation-from-rationality dynamical
model for which each agent's strategy adjustment follows the gradient of the potential along with a 
normally distributed random perturbation, is shown to equilibrate to a Gibbs measure.
The standard Cournot model of an oligopoly is shown not to have a phase transition,
as it is equivalent to a continuum version of the Curie-Weiss model.
However, when there is increased local competition among agents, a phase transition
will likely occur.  If the oligopolistic competition has power-law falloff
and there is increased local competition among agents, then the model has a rich phase
diagram with an antiferromagnetic checkerboard state, striped states and maze-like states
with varying widths, and finally a paramagnetic state.
Such phases have economic implications as to how agents compete 
given various restrictions on how goods are distributed.  The standard Cournot model
corresponds to a uniform distribution of goods, whereas the power-law variations
correspond to goods for which the distribution is more localized.
\end{abstract}

\maketitle


\newtheorem{lemma}{Lemma}
\newtheorem{theorem}[lemma]{Theorem}
\newtheorem{proposition}[lemma]{Proposition}
\newtheorem{corollary}[lemma]{Corollary}
\newtheorem{axiom}[lemma]{Axiom}

\theoremstyle{remark}
\newtheorem*{definition}{Definition}
\newtheorem*{remark}{Remark}
\newtheorem*{remarks}{Remarks}
\newtheorem{example}{Example}
\newtheorem*{ack}{Acknowledgments}


\newcommand{\gtApproach}{{Gibbs }}
\newcommand{\dfr}{{deviations-from-rationality }}

\newcommand{\nt}{\negthickspace}
\newcommand{\ntfive}{\nt \nt \nt \nt \nt}
\newcommand{\nttwenty}{\ntfive \ntfive \ntfive \ntfive}

\newcommand{\unitx}{\mathbf{\hat{x}}}
\newcommand{\unity}{\mathbf{\hat{y}}}
\newcommand{\unitz}{\mathbf{\hat{z}}}
\newcommand{\con}[1]{\overleftrightarrow{_{#1}} } 

\newcommand{\lemRef}[1]{Lemma~\ref{#1}}
\newcommand{\thmRef}[1]{Theorem~\ref{#1}}
\newcommand{\propRef}[1]{Proposition~\ref{#1}}
\newcommand{\corRef}[1]{Corollary~\ref{#1}}
\newcommand{\exRef}[1]{Example~\ref{#1}}
\newcommand{\remRef}[1]{Remark~\ref{#1}}
\newcommand{\axRef}[1]{Axiom~\ref{#1}}
\newcommand{\eqRef}[1]{equation~\eqref{#1}}
\newcommand{\primeEqRef}[1]{equation~\textup{(\ref{#1}$'$)}}
\newcommand{\primeeqref}[1]{\textup{(\ref{#1}$'$)}}
\newcommand{\secRef}[1]{Section~\ref{#1}}
\newcommand{\appRef}[1]{Appendix~\ref{#1}}
%
%
%
%
\newcommand{\np}{{n}}					
\newcommand{\vol}{\Lambda}				

\newcommand{\vty}{\chi}				

\newcommand{\strat}{\mathfrak{s}}
\newcommand{\strats}{S}

\newcommand{\vx}{\vec{x}}
\newcommand{\vw}{\vec{w}}
\newcommand{\vy}{\vec{y}}
\newcommand{\grad}{\vec{\nabla}}

\newcommand{\vq}[1][{\np}]{(q_1,\dots,q_{#1})}

\newcommand{\eval}[1]{|_{#1}}			

\newcommand{\btt}{\tilde{\beta_t}}		
\newcommand{\bto}{\tilde{\beta_0}}

\newcommand{\br}[1]{\langle {#1} \rangle}			
\newcommand{\brf}[1]{\langle {#1} \rangle^{\wedg}}	
\newcommand{\Br}[1]{\left\langle {#1} \right\rangle}	
\newcommand{\brn}[2][{}]{\langle {#2} \rangle_{n_{#1}} }	
\newcommand{\brfn}[2][{}]{\langle {#2} \rangle^{\wedge}_{n_{#1}} }
\newcommand{\brh}[1]{\langle {#1} \rangle_h}
\newcommand{\Brh}[1]{\left\langle {#1} \right\rangle_h}
\newcommand{\brnh}[1]{\langle {#1} \rangle_{n,h}}

\newcommand{\qbr}[1]{[\![ {#1} ]\!]} 				
\newcommand{\Qbr}[1]{\left[\!\left[ {#1} \right]\!\right]}

\newcommand{\qmin}{{\underline{q}}}
\newcommand{\qmax}{{\bar{q}}}

\newcommand{\adj}[1]{ {#1}^{\!\text{*}} }

\newcommand{\e}{\operatorname{e}}
\newcommand{\E}{\operatorname{E}}
\newcommand{\C}{\operatorname{C}}
\newcommand{\F}{\operatorname{F}}
\newcommand{\Arg}{\operatorname{Arg}}

\newcommand{\sinc}{\operatorname{sinc}}
\newcommand{\sinhc}{\operatorname{sinhc}}

\newcommand{\sgn}{\operatorname{sgn}}
\newcommand{\argmax}{\operatorname{argmax}}

\newcommand{\Prob}{\operatorname{Prob}}

\newcommand{\edge}{e}					
\newcommand{\bond}[1][ij]{\edge_{#1}}		
\newcommand{\bondConfig}[1][{}]{\omega_{#1}}	

\newcommand{\cfg}{\underline{\sigma}}		

%
%
\section{Introduction} \label{S:Introduction}
	Certain models in game theory \cite{AGH1,AGH2,BS,BSV,HSa1,HSa2,MaZ,MaZ2,Sa1,Sa2} analyze the dynamics of decisions made by agents who adjust their decisions in the direction of higher payoffs, subject to random error and/or information (deductive ``\dfr\nt'' models, cf. \cite{Bl}).  These errors, which are essentially failures to choose the most optimal payoff, are understood to be intrinsic to the agents, and can be due to stochastic elements such as preference shocks, experimentation, or actual mistakes in judgment.  In all of these contexts, the error is assumed to be due to intrinsic properties of the agents; i.e., the error is due to the \emph{agents} making decisions that deviate from the true, optimal decision.  These stochastic approaches commonly use a Langevin equation to deduce logit equilibrium measures for each individual agent.  

	The logit measures had been earlier discovered by different reasoning \cite{L, McF}.  Logit measures are exploited in \cite{AGH1, AGH2}, where agents use a continuum of pure strategies with decisions perturbed by white noise, and additionally, agents can use a restricted form of mixed strategies which are absolutely continuous with respect to the Lebesgue measure.  A purely dynamical approach is taken in \cite{MaZ, MaZ2}, where a continuum of pure strategies is assumed and the agents are restricted to using only these pure strategies with their decisions perturbed by white noise.  If agents are restricted to play a single pure strategy at any given point in time as in \cite{MaZ, MaZ2}, mixed strategies in \cite{AGH1, AGH2} can be viewed as a time-average of the agent's pure strategies.  Since averages are effectively used in the stationarity equations of \cite{AGH1, AGH2}, it is not surprising that the solutions for the logit measures are mean-field theoretic in nature insofar as they must find fixed points for these equations to solve them.  
A mean-field static equilibrium is also found for a discrete Ising-type logit model in \cite{BlD}.
The dynamics of a random utility model are also analyzed in \cite{BlD}, and shown to have the same
stationary state as with the mean-field analysis.  The static equilibrium is that for a
Curie-Weiss model.  Such concepts will be examined below within the context of statistical
mechanics. 
For the dynamics presented in this paper, we will assume that agents are \emph{myopic} in pure strategy space; that is they play a single pure strategy at any point in time as in \cite{MaZ, MaZ2}, and can only make infinitesimal adjustments in pure strategy space over infinitesimal time.  The difference between the dynamical approach in this paper versus those in \cite{AGH1, AGH2, Bl, 
HSa1, HSa2, MaZ, MaZ2} is that a global approach is used
	\footnote{A single dynamical equation tracks all agents, instead of one equation 
		    for each agent.}, 
in full exploitation of potential games \cite{MS}.  As a result, a single stationary measure results, instead of multiple coupled measures in \cite{AGH1,AGH2}.  An explicit form of the measure
is also obtained in contrast to the aforementioned references.  

It is noted that to allow agents to make effectively large moves in pure strategy over infinitesimal time (effectively what happens in \cite{AGH1, AGH2}) is akin to allowing mixed strategies, and as a result, Nash equilibria can be found much more quickly.  As an example, consider pure-strategy quantities $q_1$, $q_2$, and $q_3$ which are positive real numbers.  The dynamics of this paper does not allow a quick jump from $q_1$ to $q_2$ since $|q_1-q_2|>0$, where $|\cdot|$ is the distance function on the line.  However, with mixed strategies, an agent can shift from $q_1$ to $(1-\epsilon)q_1 + \epsilon q_2$ for small epsilon, and get a quick sense of the strategy $q_2$ from $q_1$ and shift towards $q_2$ in one step if it is better.  This can happen since $\|q_1 - [(1-\epsilon)q_1 + \epsilon q_2]\|<2\epsilon$, where $\|\cdot\|$ is the distance function on the Banach space of Radon measures.  This is similar to classical, noiseless potential game theory when the Nash equilibrium is quickly and efficiently found to be a pure state (i.e., the global interior maximum of the potential function) since agents can make large unilateral changes in decision.  This does not happen in the myopic, noisy dynamics presented here at nonzero temperature, as is evident from the resulting mixed strategy Gibbs/logit stationary state.  The classical game-theory equilibrium (i.e., Nash equilibrium)
is attained from the Gibbs measure at zero temperature.  A thorough analysis of types of equilibria
that occur in potential games can be found in \cite{Sa1}.

	In addition, an alternative view is proposed in this paper, in the context of potential games.  The same Gibbs/logit equilibrium distribution that results from myopic, noisy dynamics can be derived using certain axioms.  This avoids explicit reference to error in agents' decisions and extends easily to games with discrete pure strategy spaces.  These axioms are in fact those for equilibrium thermodynamics and are used to do large deviation theory.  Therefore this alternative approach is empirical in nature: our model of agent behavior using thermodynamic axioms is based on many observations rather than the intrinsics (i.e. dynamics) of the game.  It will be shown that this \emph{\gtApproach\nt} approach to games with a continuum of strategies is equivalent to the aforementioned global, noisy, myopic \dfr potential game.  As in classical statistical mechanics, we will assume the observables (i.e., functions of strategies) to be $C(\Omega)$, the continuous functions on pure strategy space.  Hence the general mixed strategies will be the probability states in the dual of $C(\Omega)$, $R_1(\Omega)= \{\mu \in C(\Omega)^* |\, \mu \geq 0, \mu(1)=1\}$.  That is to say the mixed strategies will be generalized to probability Radon measures on $\Omega$, which can include density functions with Lebesgue measure (such as the Gibbs measure), and delta functions for pure states (zero-temperature/pure-rational limits of Gibbs measures).

Since we only use one dynamical equation, only one stationary logit measure is obtained and we will call this measure the \emph{Gibbs measure} of the system, exactly as in statistical mechanics.  This will also avoid confusion with the more general case presented in \cite{AGH1, AGH2} of multiple logits for each agent.  The Gibbs measure here is one for the entire system of agents in the given game which results from the global dynamical equation.  This is a very interesting feature of potential games with noisy, myopic dynamics.  It is stated in \cite{MaZ2} that ``...in game theory, there is not a unique Hamiltonian, but rather each agent has her own Hamiltonian to minimize.''  This is true in general, but we'll show that we do have a unique Hamiltonian in the case of potential games.  The idea of looking for something global to unify dynamics was touched upon in \cite{BCP} with the use of a ``global cost function'' in the `minority game', but the technique is through the use of another dynamical method called ``inductive reasoning'' \cite{A}, rather than errors to deductive reasoning.  Within inductive reasoning, an explicit formula for the stationary measure in terms of the state of the agents does not appear.  The equilibrium measures obtained here do have such an explicit formula: a Gibbs measure in terms of the potential.  The potential was pointed out in \cite{AGH1} to be the common part of the payoff functions that includes the benefits or costs that are determined by each agent's own decisions (e.g., the effort costs).  Mathematically, the potential is just like the negative energy of all of the particles (i.e. agents) in a physical system (i.e. game) for a given state (i.e. point describing each agent's decision).

	The essence of the thermodynamic approach is that we are measuring many cases of the same game, as one would repeat an experiment many times to get average results; i.e., an \emph{ensemble average}.  The errors are not explicitly encoded at the (microscopic) level of individual agent decisions in this approach.  However, this endogenous noise is implicitly assumed on a macroscopic level by requiring the ensemble average of the potential be constant.
	\footnote{If one game in the ensemble gains potential, another must 
	lose potential which is counter to purely rational behavior.}
The average values are due to making many measurements and averaging the results over the \emph{ensemble} of the multitude of games being observed.  Hence it's assumed that we don't know the details of probabilities of being in various states, but rather that we must deduce these probabilities from axioms of empirical observation.  The size of the ensemble (i.e., the number of representative games we observe) must be very large so that statistical laws of large numbers hold, and that the empirical probabilities reflect the true intrinsic probability distribution of a single game.  However, unlike the ideas proposed in \cite{Ma, MaZ, MaZ2}, we do not need to assume that there are a large number of agents and we do not treat a single agent as though she is a system in a `heat bath' of agents, which implies an ensemble of agents and that we are only looking at one game.  Rather, we treat the entire game as a system, and look at an ensemble of many identical games.  This approach also allows an easy generalization to discrete/finite decisions.  The partition function is introduced systematically and axiomatically, rather than as a post-hoc mechanism to solve a minimum free energy problem in inductive models as in \cite{CGGS, MaCZ}.  The logit measure follows very naturally from the framework presented here.  The limitation is that, for now, a systematic introduction of a partition function (i.e., explicit stationary measure) is only justified for potential games, whereas \cite{AGH1, AGH2, MaCZ} is more general since agents have individual measures.

	An interesting aspect of this \gtApproach approach is that the coefficient that determines the importance of errors in a stochastic potential game (the $\mu_i=2/\sigma^2$ in \cite {AGH1, AGH2} or the $2/\nu^2$ in \eqRef{E:FokkerPlanck}) is directly proportional to the usual notion of temperature in statistical mechanics.  This was also discussed in \cite{Ma} for individual agents.  In fact, in the kinetic theory of gases in equilibrium, the temperature $T$ is proportional to the average of the square of velocities $v$ of the gas particles ($T = k\cdot\br{v^2}$ for a constant $k$).   If a game has a potential that is the mathematical equivalent of such a gas, this equilibrium `kinetic' theory states that the `temperature' of that game is proportional to the average of the square of the temporal rates-of-change of decisions ($T = c\cdot\br{[dx/dt]^2}$, where $x\in[\underline{x},\bar{x}]$ represents an agent's decision).  Hence larger instantaneous changes in decisions by agents correspond to higher temperatures of the game, and in general, temperature is a measure of the rate-of-change in decisions.  In the dynamical sense \cite{AGH1, AGH2}, temperature measures agents' deviation from rationality.

	A major application of statistical mechanics is to take infinite-agent limits from finite agent games, and an example of this will be done to show the classical result of oligopolies approaching perfect competition as the number of agents goes to infinity.  This requires some modest technical restrictions to be made on the potential function so that the finite-agent Gibbs measures will converge in the infinite-agent limit (c.f. \cite{Si}); however these restrictions also make clear economic sense.  A major advantage with the \gtApproach approach for potential games is that, for various potentials, all of the rigorous techniques for phase transitions can be implemented as well as the process of renormalization of infinite-agent games.  Renormalization would allow us to look at behavior of aggregate blocks of agents at various scales. Dynamical renormalization is also very recently being done in economics \cite{FJL, FL}, where large numbers of small firms are ``agglomerated'' and the dynamics of the resulting ``meta-firms'' are examined via a sandpile model. It will only be mentioned in this paper that renormalization can be done.  The standard real-space renormalization is applicable to any potential game with the imposition of a lattice dimension, since all agents interact in many classic economic models.  It would also be interesting to compare dynamical renormalization of the noisy myopic dynamics with the real-space renormalization of the Gibbs measure.  Hopefully, the model shown here may be fruitful in understanding how local and global interactions between agents unify; that is to say a better understanding and more systematic treatment of economic interactions at small and large scales.  Renormalization may link different economic scales as it does different scales in physics.

%
%
\section{Potential Games and Axioms} \label{S:pGamesAxioms}

	Let us consider a game with a finite number of $\np$ agents, and all of these agents belong to the set $\vol$.  At any moment in time, agent $i\in\vol$ can select an action or strategy $x_i \in A_i$ and the $x_i$ are the \emph{decision variables}.  A \emph{configuration} $\vx$ of the system is any possible state of the system:  $\vx = (x_1, x_2, \dots, x_\np)$, where each $x_i \in A_i$.  The set of all possible configurations of the game is $\Omega \equiv \prod_i A_i$.  A typical example of $A_i$ is an interval of real numbers $A_i=[\underline{x},\bar{x}]$.  The development below is for continuum decisions, and discrete decisions will be accommodated later. 

In this article, we will only consider potential games \cite{MS}.  Let
	\begin{equation}
 	V(\vx)=V(x_1,x_2,\dots,x_\np):\Omega\rightarrow(-\infty,\infty)
	\label{E:potential}
	\end{equation}
be the potential function for the game.  Recall that a potential game is one such that every agent's payoff function $u_i(\vx)$ satisfies
	\begin{equation}
	\frac{\partial u_i}{\partial x_i} = \frac{\partial V}{\partial x_i}
	\label{E:potentialCond}
	\end{equation}
for some function $V(\vx)$.

Certain conditions must be imposed upon $V$ in order that probability distributions exist when the number of agents is taken to infinity in a limit.  These conditions are well known functional-analytic restrictions to ensure that infinite-volume limits exist (c.f. \cite{Si}).  Basic postulates for a game in equilibrium will be introduced (c.f. \cite{R}).  These are just the usual thermodynamics, and the probability of finding the game in a particular state can be derived from these postulates of empirical observation.

\begin{axiom}[conservation of energy] \label{A:axiom1}
The game is \emph{isolated} and agents act in such a way that the ensemble average of the potential is a constant.\end{axiom}

The `isolated' requirement excludes outside sources from changing the potential for a given set of choices.  Even though agents in a specific game in the ensemble try to maximize potential, the intrinsic noise will introduce errors into agents' decisions such that an increase in potential in one game will correspond to a decrease in the potential in one or more other games in the ensemble.  Since purely rational agents attempt to maximize potential, this axiom precludes perfect rationality unless the ensemble average of the potential \emph{is} the maximum potential (this is the zero-temperature case).

\begin{axiom}[equilibrium/stationarity] \label{A:axiom2}
The isolated game is in \emph{equilibrium}.  This is to say that the probability of finding the game in any one state (i.e., a single representative from the ensemble) is independent of time.  
\end{axiom}

This is the usual commonsense notion of equilibrium.  From \axRef{A:axiom2}, all \emph{macroscopic} parameters (those that describe the system as a whole) are then also time-independent.  Macroscopic parameters will be described in more detail later, but examples are the average output per agent, the average payoff, and the temperature.

\begin{axiom}[ergodicity] \label{A:axiom3}
The isolated game in equilibrium is equally likely to be in any of its accessible states.  
\end{axiom}

An \emph{accessible state} $\vx$ is any configuration of decisions $\vx\in\Omega$ with $V(\vx)=V_0$, given a fixed observed potential $V_0$ for the game (i.e., we look at all games in the ensemble with potentials of $V_0$).  If a potential game with potential $V$ has payoff functions $u_i(\vx)$ for each agent $i\in\vol$, then
	\begin{equation}
	\frac{\partial}{\partial x_i} u_i(\vx) = 
		\frac{\partial}{\partial x_i} V(\vx).
	\label{E:gradV}
	\end{equation}
If the potential for $\np$ agents is restricted to a constant value $V_0$ (i.e., we look at a constant-potential surface in $\Omega$), then the equilibrium dynamical system of \cite{AGH1} is ergodic.  Fluctuations will cause a system to move among its accessible states over time with equal frequency, and ergodicity means essentially that the system, over time, will visit each point on the constant potential surface.  As such, it is equivalent to assume all points on the constant $V_0$ surface have the same probability of being observed.  An interesting consequence of this that the condition $\partial V/\partial t =0$ does not necessarily imply an equilibrium (c.f. \cite{AGH1}).  If the system were at a local maximum of $V$ (at say $\vx_0$) that isn't a global maximum, then \axRef{A:axiom3} implies the system can't stay at the single point $\vx_0$.  Rather, over time, it will move on the whole surface $V=V(\vx_0)$.  Since $V(\vx_0)$ isn't a global maximum, the system will reach another point with the same potential but with nonzero gradient, and the dynamics can move the system towards a higher potential than $V(\vx_0)$.

	The approach to stochastic game theory in \cite{AGH1} produces a single Fokker-Planck equation that the joint probability distribution of decisions on $\Omega$ satisfies.  There, the Langevin equations for the joint distribution are
	\begin{equation}
	dx_i(t) = \frac{\partial}{\partial x_i} u_i(\vx,t)dt + \nu \, dw_i(t),
		\: 1\leq i\leq\np,
	\label{E:Langevin}
	\end{equation}
where the $w_i$ are zero-mean, unit-variance normal random variables and $\nu$ is a variance parameter.

The following result very similar to \cite{AGH1}.  The major difference here is that we are looking at a dynamics for all agents who only use pure strategies, whereas in \cite{AGH1} single-agent dynamics are analyzed separately, and they use mixed strategies.  The point of the following result is to show the stochastic dynamical approach in global form leads to the same Gibbs measure as that derived from \axRef{A:axiom1}, \axRef{A:axiom2}, and \axRef{A:axiom3}.  The proof is in \appRef{S:appendixA}.  Again, it is important to note that the dynamics below assume agents only use pure strategies and can only make small-distance moves over pure strategy space within a small time period (myopic agents).

%
\begin{proposition} \label{P:potentialGame}
Let $f(\vx)$ be the joint density over decision space $\Omega$ for a potential game with a finite number of agents $\np$ and potential $V$.  Consider the dynamics
	\begin{equation}
	d\vx = \grad{V}dt + \nu d\vw(t),
	\label{E:langevin}
	\end{equation}
where $x\in\Omega$, $d\vx=(dx_1,\dots,dx_\np)$, $\grad{V}=(\partial V/\partial x_1, \dots, \partial V/\partial x_\np)$, and $\vw(t)=(w_1(t),\dots,w_\np(t))$ with the $w_i(t)$ standard Wiener (or white noise) processes which are identical and independent across agents and time.  Furthermore, the $w_i(t)$ have mean zero and variance one and reflecting boundary conditions
	\footnote{This requires zero time derivatives on the boundary,
	specifically that \eqref{E:statState} be satisfied for boundary points $\vx$.}
are used.  Note that no conditional averages are done on $V$ above, as opposed to \cite{AGH1, AGH2}.  This indicates that only pure strategies are being played.

If the process $\vx(t)$ satisfies the dynamics of \eqref{E:Langevin}, then the joint density satisfies the Fokker-Planck equation
	\begin{equation}
	\frac{\partial f(x,t)}{\partial t} =
	  -\grad\cdot[\grad{V}(\vx(t)) f(\vx,t)] + \frac{\nu^2}{2} \nabla^2 f(\vx,t)
	\label{E:FokkerPlanck}
	\end{equation}
and the corresponding equilibrium measure for variance $\nu^2$ is the Gibbs state
	\begin{equation}
	f(\vx,t) = f(\vx) =\frac{exp\left( \frac{2}{\nu^2} V(\vx) \right)}
				{\int_\Omega exp\left( \frac{2}{\nu^2} V(\vy) \right) d\vy}.
	\label{E:Gibbs}
	\end{equation}
\end{proposition}

In statistical mechanics, the term in the exponent of \eqref{E:Gibbs} is $-E(\vx)/(kT)$, where $k$ is Boltzmann's constant, $T$ is temperature, and $E(\vx)$ is the energy of configuration $\vx$.  Hence the analogy of a potential game to statistical mechanics is that $\nu^2$ (deviation from rationality; influence of the noise in dynamics \eqref{E:Langevin}) is proportional to `temperature' and the potential $V$ is the negative `energy' of the system.

Now the Gibbs measure above can be derived using only the axioms of thermodynamics \axRef{A:axiom1}, \axRef{A:axiom2}, and \axRef{A:axiom3}.  The proof is the standard proof of the `canonical distribution' in any text on thermodynamics, and is the one that finds a state (i.e. probability measure) that maximizes entropy under the condition of a constant ensemble-average energy.  The variational form of this maximization problem is the Gibbs variational principle of finding a state that maximizes the Helmholtz free energy.  It is interesting to note that the problem of finding a state that maximizes the Liapunov function (for dynamical equilibrium) in \cite{AGH1} is identical to finding a state that minimizes the Helmholtz free energy of a given potential game.  This is exactly what the Gibbs variational principle accomplishes.  It is no surprise that the Liapunov function in \cite{AGH1} \emph{is} the negative Helmholtz free energy for corresponding potential game with no explicit noise.  This fact is stated here as a contrast to \propRef{P:potentialGame}:

%
\begin{proposition} \label{P:potentialGameAxiom}
Let $f(\vx)$ be the joint density over decision space $\Omega$ for a potential game with a finite number of agents $\np$ and a potential $V$.  Suppose  the three thermodynamic axioms \axRef{A:axiom1}, \axRef{A:axiom2}, and \axRef{A:axiom3} hold.  Then the equilibrium measure for this potential game is the Gibbs state at inverse-temperature $\beta\equiv(kT)^{-1}$:
	\begin{equation}
	f(\vx,t) = f(\vx) =\frac{exp\left( \beta V(\vx) \right)}
				{\int_\Omega exp\left( \beta V(\vy) \right) d\vy}.
	\label{E:Gibbs2}
	\end{equation}
\end{proposition}

Here the constant $\beta$ in $f(\vx)$ is determined so that the mean potential $\int_\Omega V(\vx) f(\vx) d\vx=\bar{V}$, where $\bar{V}$ is the ensemble-average observed potential (i.e., the sum of the potentials of each game in the ensemble divided by the number of games $g$ in the ensemble $(V_1+\cdots+V_g)/g$).  This is equivalent to the notion of considering one game to reside within the `heat-bath' of the other games in the ensemble with a fixed total ensemble potential $V_1+\cdots+V_g$.

We consider equilibrium states of an infinite-agent game to be the measure(s) resulting from the appropriate infinite-agent/volume limit of the Gibbs state.  

In the case of potentials that fit the usual notion of an `interaction'
	\footnote{It is noted that the potential for a Cournot oligopoly
	is \emph{not} such an `interaction', and other techniques must be
	used.  This is because of the $1/\np$ factor; see \secRef{S:Cournot}
	for the explicit potential.} 
in statistical mechanics (c.f. \cite{Si} for the appropriate machinery), the infinite-agent measures will be consistent with the finite-agent measures via the Dobrushin-Lanford-Ruelle equations.  Such techniques can be used directly to analyze possible phase transitions for an infinite-agent potential games.  We can then consider equilibrium states of an infinite-agent game to be the appropriate infinite-volume limit of the Gibbs measure.

%
%
%
%
%
%

\section{Example: Cournot Oligopoly, Local Demand, Perfect Competition and Collusion} \label{S:Cournot}
The simplest oligopoly model is due to Augustin Cournot (1838).  There is one homogeneous good with demand function $p(Q)$, where $Q$ is the total quantity of the good produced.  Given an oligopoly of $\np$ firms, if each firm produces an amount $q_i$ of the good, then
	\begin{equation}
	Q = \sum_1^{\np} q_i.
	\label{E:Q}
	\end{equation}
For the sake of later analysis, each firm's production $q_i$ will be scaled between a minimum production $\qmin\geq0$ and a maximum production $\qmax>\qmin$.  We will assume each firm produces a sufficiently large number of the good so that the $q_i$ can be regarded as continuum variables $\qmin\leq q_i\leq\qmax$, $1\leq i\leq \np$.  For example, $q_i=1$ can represent the production of a sufficiently large number of goods.  It is noted that smaller production would be handled in the discrete case as in \secRef{S:MG}.   

Each firm uses quantity $q_i$ as its strategy based on the payoff function
	\begin{equation}
	\Pi_i = q_i p(Q) - C_i(q_i),
	\label{E:generalPayoff}
	\end{equation}
where $C_i$ is the $i$th firm's cost function.  We will assume a linear (inverse) demand function
	\begin{equation}
	p(Q) = a - \frac{b}{\np} Q,
	\label{E:linearDemand}
	\end{equation}
with constants $a>0$ and $b>0$.  Notice that $b$ is divided by $n$ so that demand is based on the average production.  Thus demand stays nonnegative for large $n$.  For example, if each firm were to produce $\qmax$, $p(n\qmax) = a - b\qmax$ is well-behaved and non-trivial (i.e., doesn't go to negative infinity or zero). 

Constant marginal costs (i.e., $C_i'=c$, for a constant $c>0$) will also be assumed, so that $C_i(q_i) = c q_i + d_i$, and the payoff functions are now   
	\begin{equation}
	\Pi_i = q_i \frac{-b}{\np}\sum_1^n q_j + (a-c)q_i - d_i.
	\label{E:payoff}
	\end{equation}

In the case of collusion, firms would try to maximize total industry profits
	\begin{equation}
	\Pi = \sum_1^{\np} \Pi_i = -\frac{b}{\np}\sum_{i,j=1}^{\np} q_i q_j 
				+ (a-c)\sum_1^{\np} q_i - \sum_1^{\np} d_i.
	\label{E:payoffC}
	\end{equation}

Both of these cases constitute potential games.  A potential for the oligopoly in \eqRef{E:payoff} is
	\begin{equation}
	V_o = -\frac{b}{2\np} \sum_{i,j} q_i q_j - \frac{b}{2\np}\sum q_i^2
		+ (a-c)\sum q_i
	\label{E:potentialO}
	\end{equation}
and a potential for collusion is the same as the collusion payoff function (except for the irrelevant constants $d_i$) in \eqRef{E:payoffC},
	\begin{equation}
	V_c = -\frac{b}{\np} \sum_{i,j} q_i q_j + (a-c)\sum q_i
	\label{E:potentialC}
	\end{equation}
These two potentials are two cases of the potential
	\begin{equation}
	V = -\frac{b}{\np} \sum_{i,j} q_i q_j - \frac{\tilde{b}}{\np}\sum q_i^2
		+ (a-c)\sum q_i.
	\label{E:potentialCO}
	\end{equation}
To get the potential in \eqRef{E:potentialO}, replace $b$ and $\tilde{b}$ in \eqRef{E:potential} with $b/2$.  To get the collusion potential in \eqRef{E:potentialC}, set $\tilde{b}$ to 
zero.
	\footnote{
	It is interesting to note the effect of collusion pertaining to rationality.
	If $a-c=0$ in \eqref{E:potentialCO}, then collusion has the effect of doubling
	$\beta$ in the partition function \eqref{E:partitionCO}.  The collusive model
	is simply the non-collusive oligopolistic model at half the temperature 
	$1/(2\beta)=T/2$.  In the infinite-agent limit, the oligopolistic model is
	the same as perfect competition.
	Hence collusion is more ``rational'' behavior than perfect competition!}

The Nash equilibrium for the Cournot oligopoly with potential $V$ above is
	\begin{equation}
	q_j^* = \frac{a-c}{2( b + \tilde{b}/\np ) }.
	\label{E:qEquil}
	\end{equation}
For non-collusion, $q_j^* = [\np/(\np+1)] [(a-c)/b]$, and for collusion $q_j^* = (a-c)/(2b)$ are the Nash equilibria.
	\footnote{These equilibria look like those for \emph{total} industry output 
		    $Q=\sum q_i$ in the standard literature, but as described below 
		    \eqref{E:linearDemand}, the scaling is at the level of firm production
		    instead of industry production (where firm production would go to zero
		    as $\np$ increases).}
Note for collusion, the output is half of that for the non-collusive model, and hence firm profit \eqref{E:payoff} is less than it would be for the non-collusive equilibrium.
These classical equilibrium solutions will be recovered in the perfectly rational case (zero temperature, $\beta=0$) below, using the \gtApproach approach.

The partition function for $\np$ agents is
	\begin{equation}
	\mathcal{Z}_{\np} = \int \prod_{i=1}^{\np} dq_i
	  \exp\left[ -\beta\frac{b}{\np} \sum q_i q_j
		-\beta\frac{\tilde{b}}{\np}\sum q_i^2 + \beta(a-c)\sum q_i \right].
	\label{E:partitionCO}
	\end{equation}

The infinite-agent free energy
	\begin{equation}
	F(\beta,a,b,\tilde{b},c,\qmax,\qmin) = 
		\lim_{\np\to\infty} \frac{1}{\beta \np} \ln( \mathcal{Z}_{\np,h} ).
	\label{E:FreeEnergyDef}
	\end{equation}
is calculated in \appRef{S:appendixB} to be
	\begin{equation}
	\begin{aligned}
	F(\beta,a,b,c,\qmin,\qmax)   
	=&\frac{ b(\qmax+\qmin)[ b(\qmax+\qmin)- 2(a-c) ]}{4\beta}
	  -\frac{\ln(\qmax-\qmin)}{\beta}
	\\
	&- \frac{1}{\beta}\min_{y\in(-\infty,\infty)}
		-\beta b y^2 
		+ \ln{\Big (} 
			\sinhc\left[ 
		        \beta b(\qmax-\qmin)y + 
			  \beta\frac{(\qmax-\qmin)}{2} \{b(\qmax+\qmin)-(a-c)\}
			\right]
		{\Big )}
	\end{aligned}
	\label{E:freeEnergy}
	\end{equation}
where $\sinhc(x)=\sinh(x)/x$, $\sinhc(0)=1$, and it is noted that the free energy $F$
is independent of $\tilde{b}$, which can be set to zero.  It is also shown in
\appRef{S:appendixB} that the expected value of any agent's output is
	\begin{equation}
	\br{q_i} = \frac{\partial}{\partial a} F 
	= \frac{a-c}{2b} + \frac{y_m(\beta,a,b,c,\qmin,\qmax)}{\beta Qb},
	\end{equation}
where $|y_m|<|(\qmax+\qmin)/2 - (a-c)/(2b)|$, $y_m<0$ when $(a-c)/(2b)>(\qmax+\qmin)/2$, 
and $y_m>0$ when $(a-c)/(2b)<(\qmax+\qmin)/2$.  Hence the Gibbs equilibrium
is always between the average $(\qmax+\qmin)/2$ and the Nash Equilibrium $(a-c)/(2b)$.
In the completely irrational limit
	\begin{equation}
	\lim_{\beta\to 0} \br{q_i} = \frac{\qmax+\qmin}{2},
	\label{E:magIrr}
	\end{equation}
and the output is pushed to the average as would be expected for 
uniformly random behavior.
In the completely rational limit,
	\begin{equation}
	\lim_{\beta\to\infty} \br{q_i}
	= \left\{
		\begin{aligned}
		&\qmin \qquad \text{if\ }&\frac{a-c}{2b}<\qmin,
		\\
		&\frac{a-c}{2b} &\qmin\leq \frac{a-c}{2b}\leq\qmax,
		\\
		&\qmax &\frac{a-c}{2b}>\qmax,
		\end{aligned}
	\right.
	\label{E:mag}
	\end{equation}
which is the classical Nash equilibrium for an infinite number of agents in \eqref{E:qEquil}.

From the properties of $y_m$, we see that agents adjust output down (up) from 
$q^*\equiv(a-c)/(2b)$ towards the average output $q_{av}\equiv(\qmax+\qmin)/2$
if $q^*$ itself is larger (smaller) than $q_{av}$.
The magnitude of the adjustment increases as the deviation from rationality 
(i.e., temperature or $1/\beta$) increases.
This shows that a collusive-type behavior occurs insofar as agents produce \emph{less} than
the purely rational equilibrium output $q^*$ when $q^*>q_{av}$.  When $q^*<q_{av}$,
agents will produce \emph{more} output than $q^*$, which is anti-collusive/supercompetitive.

It is also shown in \appRef{S:appendixB} that the \emph{volatility}, which is
$\vty \equiv\beta^{-1} \partial^2 F/\partial h^2$, is zero at zero temperature
(as expected for perfect rationality) and goes to $(\qmax-\qmin)^2/12$ as temperature goes
to infinity (i.e., randomly acting agents have independent, uniformly distributed 
outputs $q_i$). 
An agent's payoff can be computed using $\br{q}$ and $\vty$.  Since
$\br{\Pi_k} = \lim_{\np\to\infty} q_k[ -b/n \sum_{j=1}^\np q_j + (a-c)]$ and
$\br{\Pi_k} = \br{\Pi_i}$, $1\leq i\leq\np$, we have
	\begin{equation}
	\br{\Pi_k} = \lim_{\np\to\infty}
		-\frac{b}{\np}\vty  + \br{q_k}(h-b\br{q_k})
	= \br{q_k}(h-b\br{q_k}).
	\end{equation}
The maximum payoff occurs at the Nash equilibrium $\br{q_k}=q^*$.  Furthermore, from
the explicit formula for $\br{q_k}$ it is evident that $\br{\Pi_k}$ is an increasing
function of $\beta$.  Payoff increases as the deviation from rationality (i.e. temperature)
decreases.

The model presented above is the standard Cournot model, which assumes 
the inverse demand is symmetric among the agents.  In reality, this is not typically the case.
For example, water prices in parts of a city with higher elevation may have a higher price
because of additional costs of pumping the water.  Likewise, competition among agents may
not be perfectly uniform.  An example of this would be firms who are more competitive against
neighboring firms than with those far away, such as a local restaurant.  It draws customers
from mostly one area, and would be in greater competition with other nearby restaurants.  The
success or failure of distant restaurants would not have any affect.  A single member
of a restaurant franchise competes with local restaurants, but colludes with other more distant
members of the parent company.
In contrast to this would be a type of regional collusion, where all of the
firms in any given locale may work in collusion with each other in an attempt to outcompete
distant firms.

To formalize these ideas, notice that an individual payoff function can be written as
the collusion potential minus all other agents' payoff functions:
	\begin{equation}
	\Pi_k = V_c - \sum_{i\not= k} \Pi_i.
	\label{E:payDiff}
	\end{equation}
This shows an agent in oligopolistic competition will offset the collusive potential by
subtracting the gains made by \emph{all} other agents.  A smaller degree of competition
would correspond to subtracting fewer of the other agents payoffs.

\begin{example}[Local Collusion Within a Global Oligopoly]\label{Ex:lc}
Suppose $\np$ agents are assumed to lie at integer points on the real line.  If agents
work collusively in disjoint groups of three, the payoffs are of the form
	\begin{equation}
	\Pi^{lc}_{k\pm 1}=\Pi^{lc}_k
	= \sum_{i=k,k\pm 1} \Pi_i
 	= \sum_{i=k,k\pm 1} \sum_{j=1}^{\np} 
		 \left[ -\frac{b}{\np} q_i q_j + (a-c) q_i \right]
	\qquad k=3m, \text{\ for integer\ }m.
	\label{E:localCollusion}
	\end{equation}
A potential is $V_{lc}=V_o - (b/\np) \sum_{k=3m,i=k\pm 1} q_i q_k$, where $V_o$ is the
oligopolistic potential in \eqref{E:potentialO}.  Although the potential is not translation
invariant, if `block outputs' $Q_k\equiv(q_{k-1}+q_k+q_{k+1})/3$, $k=3m$, are used 
along with periodic boundary conditions ($q_\np\equiv q_1$) then
a renormalized potential would be translation invariant.
\end{example}

\begin{example}[Local Oligopoly Within Global Collusion] \label{Ex:lo}
Suppose now that agents collude with non-neighboring agents and compete oligopolistically
with neighboring agents within the same setting as \exRef{Ex:lc}.  Here, however, the
agents are competing oligopolistically with nearest neighbors, and not in disjoint groups
of three.  The payoffs are
	\begin{equation}
	\Pi^{lo}_k
	= \sum_{i\not= k\pm 1} \Pi_i= V_c - \Pi_{k-1} - \Pi_{k+1},
	\label{E:localCPay}
	\end{equation}
where we define $\Pi_0\equiv 0$ and $\Pi_{\np+1}\equiv 0$ to get correct payoffs. 
The potential is 
	\begin{align}
	V_{lo} &= V_c + \frac{b}{n}\sum_{i=1}^{\np-1} q_i q_{i+1}\\
	&=
	-\frac{b}{\np} \sum q_i q_j + (a-c)\sum q_i
		  + \frac{b}{\np} \sum_{i=1}^{\np-1} q_i q_{i+1}.
	\label{E:localOPot}
	\end{align} 
It is shown in \appRef{S:appendixB} that the $\tilde{b}$ term in the potential $V$
of \eqref{E:potentialCO} has no significance in the infinite-agent limit because
the number of terms in the summand is of order $n$, and as such it can be set to zero.
This is the case for the last nearest-neighbor sum in \eqref{E:localCollusion}, and in the
infinite-agent limit, the free energy is the same as the collusion model.
To maintain local oligopolistic competition in the infinite-agent limit, the interaction
term $b/\np$ must be changed to a constant $\delta>0$ that does not vanish as $\np\to\infty$. 
\end{example}

\begin{example}[True Local Oligopoly Within Global Collusion] \label{Ex:loReal}
As mentioned in \exRef{Ex:lo}, an appropriate potential that maintains local competition
in the infinite-agent limit is	
	\begin{equation}
	V_{lo} =
	-\frac{b}{\np} \sum q_i q_j + (a-c)\sum q_i
		  + \delta \sum_{i=1}^{\np-1} q_i q_{i+1},
	\label{E:loReal}
	\end{equation} 
where $\delta>0$.  
\end{example}

The $\delta$ term in \eqref{E:loReal} only involves \emph{local interactions}, that is,
only terms within a fixed distance (of one) appear in the summand.  Such terms in the
potential are also called \emph{local interactions}.

\begin{example}[Oligopoly With Stronger Local Competition]\label{Ex:locComp}
It is noted that in the infinite-agent limit, replacing $b$ with $b/2$ in \eqref{E:loReal} is
equivalent to changing the collusive potential component $V_c$ to the standard oligopoly
potential $V_o$.  We can then interpret the (infinite-agent) free energy generated by the
potential $V_{lo}$ to be oligopolistic competition with increased local competition.
\end{example}

The potential in \eqref{E:loReal} is also generated by payoff functions
	\begin{equation}
	\hat{\Pi}^{lo}_k = q_k \left( 
	  -\frac{2b}{\np}\sum_j q_j + (a-c) 
		+\delta \left[ q_{k-1} +\frac{b}{\delta\np}q_k + q_{k+1} \right] ,
	\right)
	\label{E:loPay}
	\end{equation}
where $q_0\equiv q_{\np+1}\equiv 0$ applies.  The dynamics generated by the
$\hat{\Pi}^{lo}_k$ in \eqref{E:loPay} and the $\Pi^{lo}_k$ in \eqref{E:localCPay}
are identical, since they generate the same potential.  Likewise, the infinite-agent
free energies are the same, and as argued previously, the term $bq_k/(\delta\np)$ will
have no effect in the infinite-agent limit.  

What is interesting about \eqref{E:loPay} is that an inverse demand function appears:
	\begin{equation}
	p_k(q_1,\dots,q_\np) = p(Q)
				+\delta\left[ q_{k-1} +\frac{b}{\delta\np}q_k + q_{k+1} \right] ,
	\label{E:locDemand}
	\end{equation}
where $p(Q)$ is a standard linear inverse demand as in \eqref{E:linearDemand}.
The remaining term only involves terms local to agent $k$, and is called a 
\emph{local (inverse) demand} function.  Also notice that payoffs cannot
be described by a single global demand $p(Q)$, and each agent has her own demand function
$p_k$.

The reason for the consideration of the addition of local demand as with \eqref{E:loReal}
is that a phase transition is possible in contrast to the lack of one for the standard
Cournot oligopoly.  This will be shown in \secRef{S:MG} for a discrete model which is
a version of the Cournot model with local demand.

Finally, we consider a situation where local demand for an agent's good is less affected
by more distant firms.

\begin{example}[Power-Law Falloff of Local Demand]\label{Ex:localDemandPower}
Imagine agents residing at integer coordinates in a box in $d$-dimensional space; call
this set $Q\subset\mathbb{Z}^d$.
Consider the local inverse demand functions
	\begin{equation}
	p_k(q_1,\dots,q_\np) = 
	- 2 b \sum_{j\in Q:j\not=k} \frac{q_j}{|k-j|^w} + (a-c) 
	+ 2 \delta\sum_{i:|i-k|=1} q_i,
	\label{E:powerDemand}
	\end{equation}
where $|i-j|$ is the distance between agents $i$ and $j$, and
$w\geq 1$.  Here, the $\delta$-term again represents heightened local competition,
but the $b$-term is in a different form.  The output of agents more distant from agent $k$
has less of an effect on the inverse demand of agent $k$.  This can be due to a variety of
issues.  Local agents may have more of a local market share because more distant agents could
have to pay more for shipping, which can increase the price of their good in more distant
regions.  A variation of that is the good is primarily bought and not shipped, with customers
preferring to travel less distance to purchase the good.  This is in contrast to the previous
term $p(Q)$ in \eqref{E:linearDemand} which assumes goods are circulated uniformly in a
global market, with no barriers that would put more emphasis on local output.

A potential for payoffs $\Pi_k=q_k p_k$ is
	\begin{equation}
	V_w =
	-b \sum_{(i,j)} \frac{q_i q_j}{|i-j|^w} + (a-c)\sum q_i
		  + \delta \sum_{\br{i,j}} q_i q_j,
	\label{E:powerPot}
	\end{equation} 
where $(i,j)$ indicates a sum over distinct pairs of sites in $Q$ and
$\br{i,j}$ indicates a sum over nearest neighbors $|i-j|=1$.
\end{example}

It has been shown that a very interesting phase transition occurs for a discrete version
of \eqref{E:powerPot} above, which will be outlined in \secRef{S:MG}.
This phase transition is very distinct from the phase transition
speculated for the discrete version with the standard $p(Q)$ term.
We then see that an interesting feature of the Gibbs model of Cournot competition is
that it distinguishes between models with uniform and nonuniform distribution of
goods for certain ranges of parameters $\beta$, $b$, and $\delta$.

%
%
%
%
%
%

\section{Example: Gibbsian Minority Game} \label{S:MG}

The inductive minority game (see, for example, \cite{CZ, MaC}) is a version of the El Farol bar
problem \cite{A, MaCZ} that was introduced to analyze cooperative behavior in markets.  It is
introduced in the context of inductive reasoning and non-equilibrium statistical mechanics, but
we will look at it in another way since the utility functions can be derived from a potential. 
An interesting point is that the dynamical equations in the inductive reasoning view do not
satisfy detailed balance and there is a dynamical phase transition separating an ergodic from a
non-ergodic process (c.f. the discussion in \cite{C}).  Similarly, there are Glauber dynamics
for the Ising model which use sequential site updating and do not in general satisfy detailed
balance, but nevertheless do converge to the appropriate Gibbs measure by stationarity (see
\cite{So} for an example).  The lack of detailed balance usually suggest a study of dynamics,
but the advantage of a \gtApproach approach is that the dynamics yield an explicit
stationary equilibrium state.  

In equilibrium analysis, many models exhibit a unique translation invariant infinite-volume
state (hence ergodicity) at high temperature, while at lower temperatures there are multiple
states and non-ergodic translation invariant limiting states exist (c.f. \cite{Si} for
details).  As such it seems reasonable to consider a \gtApproach model of the minority game. 
First, for the sake of background, we will introduce the inductive reasoning form of the game
as presented in \cite{C}, and then contrast it with the Gibbsian version.

The inductive game consists of $\np$ agents, labeled $i=1,\dots,\np$.  For simplicity, we can
picture these agents lined up at integers on a one-dimensional axis, hence the agents are
situated at sites $\vol\subset\mathbb{Z}$.
		\footnote{
		Because of the form of the potential, this model is not a $d$-dimensional
		lattice model for any $d$.  We have chosen to use $d=1$ for visualization, and
		later, a true $d$-dimensional term will be added to the potential.
		}	
At each discrete time step $t$, each agent $i$ must make a decision $\sigma_i\in\{-1,1\}$ (such
as to `buy' or `sell').  The agents decide to buy or sell based on public information $I(t)$
that is available to all agents at time $t$ (c.f. the discussion in \cite{C}).  Note that at a
given time, all agents receive the identical piece of information.  Examples of such
information could be the state of the market (decisions off all agents, the decisions of agents
on odd sites, the block/regional-averages of decisions on blocks of length $\np/100$, etc.),
the weather forecasts, change in government regulations, etc.

Profit, or utility, is made at time step $t$ by agents according to the formula
	\begin{equation}
	u_i(t) = \frac{-b}{\np}\sigma_i(t) \sum_{i=1}^\np \sigma_j(t)
	\label{E:utility}
	\end{equation}
with $b>0$.  Because the coefficient $-b<0$, the profit $u_i(t)\geq 0$ when 
$\sigma_i(t)=-\sgn\left[\sum_{i=1}^{\np} \sigma_j(t)\right]$.  The justification is that in a
market, if the majority is selling it is profitable to buy and vice-versa.  Further notice the
payoff function is scaled by $1/\np$.  It is mentioned in \cite{C} to exist for mathematical
convenience, but we will see later that it is needed for stability of the thermodynamic limit. 
Economically it makes sense to scale this way so that the inherent inverse demand function
stays finite (see \secRef{S:Cournot}).

Now the element of public information will be introduced.  It is noted in \cite{MaC} that a
phase transition occurs without the element of information, and all that is needed is
stochastic mixed strategies.  In \cite{C}, information is randomly drawn at time $t$ from a set
$\{I_1,\dots,I_p\}$.  For example, at time $t$ a number $\mu(t)\in\{1,\dots,p\}$ is drawn
randomly and the information $I_{\mu(t)}$ is given to all agents, so that $I(t)=I_{\mu(t)}$
with $1\leq\mu(t)\leq p$.  

The information $I(t)$ is converted into a configuration of $\np$ trading decisions
$\cfg(t)=(\sigma_1(t),\dots,\sigma_\np(t))$.  These are determined using `response strategies',
which are look-up tables or functions
$\mathfrak{R}_{i,\strat}=(R^1_{i,\strat},\dots,R^p_{i,\strat})\in\{-1,1\}^p$. 
Each agent $i$ has $\strats$ decision making strategies $\mathfrak{R}_{i,\strat}$
with $\strat\in\{1,\dots,\strats\}$.  The strategy that each agent $i$ plays depends
on the public information $I(t)$.  If $I(t)=I_{\mu(t)}$, agent $i$ will choose a strategy
$\strat$ and then lookup the decision $R^{\mu(t)}_{i,\strat}=\pm1$ in the vector
$\mathfrak{R}_{i,\strat}$.  Agent $i$ will then make the decision
$\sigma_i(t)=R^{\mu(t)}_{i,\strat}$, which depends on her choice of strategy 
\emph{and} public information.

Finally, the last ingredient is for agents to evaluate strategies based on the market
performance of the strategies.  At time $t$, strategy $\mathfrak{R}_{i,\strat}$ is 
profitable if and only if 
$R^{\mu(t)}_{i,\strat}=-\sgn\left[\sum_{i=1}^{\np} \sigma_j(t)\right]$.  Each agent $i$
will then measure the cumulative performance of each strategy $\mathfrak{R}_{i,\strat}$,
$1\leq\strat\leq\strats$ by a profit-indicator $p_{i,\strat}$ which is updated
at each time step $t$ by
	\begin{equation}
	p_{i,\strat}(t+1)=p_{i,\strat}(t) - 
	  \frac{b}{\np} R^{\mu(t)}_{i,\strat} \sum_j \sigma_j(t).
	\label{E:strategyEval}
	\end{equation}
Note that at time $t$, all strategies are updated for each agent.  Likewise note that the above
is an evaluation scheme of strategies, and the strategies $R$ are coupled to the actual
decision $\sigma_i(t)$ (since $\sigma_i(t)$ appears in the sum).  Additionally, these
hypothetical decisions are being tracked in time, even though some aren't actually made.  At
time $t$, each agent $i$ now chooses strategy $\strat_i(t)$ which has the best
cumulative performance $p_{i,\strat}(t)$ up to that time.

Given the external information ${\mu(t)}$, the dynamics of the minority game are given by
	\begin{equation}
	\begin{aligned}
	p_{i,\strat}(t+1) &= p_{i,\strat}(t) - 
	  \frac{b}{\np} R^{\mu(t)}_{i,\strat} A(t),
	\\
	A(t) &= \sum_{j} R^{\mu(t)}_{j,\strat},
	\\
	\strat_i(t) &= \argmax_{ \strat \in \{1,\dots,\strats\} }
					p_{i,\strat}(t).
 	\end{aligned}
	\label{E:inductiveMG}
	\end{equation} 

Some interesting observations about the above approach need be made.  Firstly, a phase
transition occurs in the above inductive model even though the model is a totally frustrated
one.  The fact that something stochastic is what introduces the phase transition makes sense
since this is what happens in the setting of spin glasses and replica symmetry breaking (c.f.
\cite{C, MaC2, BCP}).  It has also been shown in \cite{LVS, PSBS} that the inductive reasoning
model seems insensitive to the types of long range interactions that may occur. 
Speculatively, it seems that the global nature of public information can outweigh the
interactions among agents, just as a global magnetic field can alter the
susceptibility/volatility in a fully frustrated model.  However, there are some fundamental
physical and economical differences between the long-range interactions presented in
\cite{LVS, PSBS}.  The economical differences were described in \secRef{S:Cournot} as local
versus global inverse demand and/or availability of goods.  The physical differences between
such interactions are significant, as will be shown below.

This motivates us to study the minority game in a different light than the inductive approach
above.  Now we will introduce the \gtApproach model of the minority based on statistical
observation of deductive reasoning and show how this model distinguishes local and global
demand.  The \gtApproach approach is not inductive: only the values of the agents' decisions
are considered.  Which inductive strategy an agent may be using is irrelevant.  In other words,
we do not distinguish inductive strategies $\mathfrak{R}^{i,\strat}$ that an agent may
be considering, but only look at her final decision $\sigma_i$ to buy or sell.

In the simplest \gtApproach model, there is no phase transition, but the infinite-volume
volatility (c.f. \appRef{S:appendixB} for the continuum version) does display nonrandom 
behavior at finite
temperature.  There are phase transitions in the \gtApproach model if local interactions among
agents are added which represent greater competition among local firms.  This will be pursued
below.

Notice that the utility functions in \eqref{E:utility} can be deduced in the sense of \eqref{E:potentialCond} from the potential
	\begin{equation}
	V(\cfg) = \frac{-b}{\np} \sum_{i,j} \sigma_i \sigma_j,
	\label{E:mgPotential}
      \end{equation}
where $b>0$ and there are $\np$ agents.  The agents can be pictured to lie on integer points of some lattice, but dimensionality of the lattice is irrelevant since all agents interact.  If short-range perturbations are added, the dimensionality of the lattice is then relevant.
In the basic model presented here, the volatility is differentiable in the global field (i.e., no phase transition; see below).  However, we will momentarily digress from the basic model in ~\eqref{E:mgPotential} and look at variations in an attempt to determine what may cause a phase transition.

Consider this new model (in \eqref{E:mgPotential} below) to lie in two dimensional space, and the addition of nearest neighbor ferromagnetic/aligning interactions.  These local aligning interactions correspond to the (local) inverse demand for an agent
	\begin{equation}
	p_i(\cfg) = \frac{-b}{\np} \sum_j \sigma_j + \delta \sum_{j:|j-i|=1} \sigma_j,
	\label{E:invDemand}
	\end{equation}
where $b>0$ and $\delta>0$.  When $\np$ is large, the local aligning interactions having coefficient $\delta$ correspond to the $i$th agent's inverse demand function increasing in response to local supply increase (i.e. selling, or the $\sigma_j$ changing from $+1$ to $-1$ for $|j-i|=1$).  On the surface, this seems contrary to usual supply and demand, since inverse demand should be a decreasing function of supply variables $\sigma_j$.
	\footnote{One interpretation in \secRef{S:Cournot} is that such aligning terms are due to
		    heightened local competition.  Aligning terms result from agents subtracting the
		    utility/payoff of nearest-neighbor agents from their own utility functions.}
Mathematically, these interactions give more probability weight in the Gibbs measure to configurations with more alignments between local agents.
This could be interpreted as a reward when nearest agents to do the same thing (i.e., align: all buy or all sell).  It could be a purely psychological reward for ``following the herd'', and not part of the monetary economic demand function, since utility or payoff functions need not be based on monetary payoff alone.  It could also be a form of monetary influence where agents are willing to pay other agents to go along with their ideas.  Since an agent only has a relatively small amount of money, they would only be able to payoff several agents.  However, it is important to
note that the first term of \eqref{E:invDemand} may penalize the agent for alignment.  Ultimately,
whether or not an agent benefits from aligning depends on the phase of the system.

A potential for the model with payoffs in \eqref{E:invDemand} is 
	\begin{equation}
	V(\cfg) = \frac{-b}{\np} \sum_{i,j} \sigma_i \sigma_j
		+ \delta \sum_{\br{i,j}} \sigma_i \sigma_j,
	\label{E:mgLongShort}
	\end{equation}
where $b>0$, $\delta>0$, $i,j\in\mathbb{Z}^2$, and the last sum is over nearest neighbors 
$|i-j|=1$.  It has been speculated that the model in \eqref{E:mgLongShort} has a transition as
$T=1/\beta\to 0$ from a fully frustrated state to a ferromagnetic, or fully aligned, state
provided $b<4\delta$ \cite{Cannas}.  If $b>4\delta$ there would likely be no transition and
only full frustration as with \eqref{E:mgPotential} \cite{Cannas}.  Completely aligned states
should occur in some cases, and there should not always be only fully frustrated states.

In the case of overhyped introductory price offerings, essentially everyone is in a frenzy to
buy and we would see the configuration of $\sigma_i=-1$ for all $i$.  If news got out that a
company lost all of its worth, everyone would sell and we would see the configuration of
$\sigma_i=+1$ for all $i$. The aligning interactions are the only way to support such states
without the addition of a field term $(a-c)\sum\sigma_i$, as the model in \eqref{E:mgPotential}
has no phase transition (cf. \appRef{S:appendixB}).  The addition of such a `field' term, which
has demand and cost components, will support such states in the absence of a phase
transition.  The demand/cost component should be able to support such states, since such extreme
market reactions result in pricing adjustments (changes in $a$).  
However, it was shown in \secRef{S:Cournot} that the aligning
terms can be interpreted as increased local competition.  Such effects can become large
in such a frenzied environment, and large alignments may occur no matter how small the
demand/cost term $a-c$ may be if the temperature is low enough (i.e., agents are not
deviating much from ``rational'' behavior).

The model in \eqref{E:mgLongShort} is similar to the models studied and summarized in 
\cite{MI,CGT,MM}. The salient difference is that \cite{MI,CGT,MM} study a long-range `Coulomb' 
($w=1$) or `dipole' ($w=3$) on the two-dimensional square lattice 
	\begin{equation}
	V(\cfg) =  -b \sum_{i,j} \frac{\sigma_i \sigma_j}{|i-j|^w}
		+ \delta \sum_{\br{i,j}} \sigma_i \sigma_j,
	\label{E:mgDipoleShort}
	\end{equation}
where the $i,j\in\mathbb{Z}^2$ are two-component vectors with integer
coordinates. These  models are very rich, and interesting in an economic sense: the local
demand functions should depend more strongly on local agents, and not depend significantly on
distant agents.  This suggests that output of goods local to an agent has more of an affect
on that agent than goods that are distant from the agent, i.e., that 
\emph{local inverse demand}
(cf. \secRef{S:Cournot}) depends more on local supply.  The model in \eqref{E:mgLongShort}
assumes that local inverse demand is identical for each agent, and that distant goods are equally accessible to an agent as local goods.  There is a fundamental difference between these
interactions.  An important difference in two dimensions between \eqref{E:mgLongShort} above
and models with falloff potentials where $\np$ is replaced with $|i-j|^w$, $w=1,3$, is that the
latter models can have a phase transition to an antiferromagnetic phase at low temperatures
and the former model has no such phase transition and only totally frustrated ground states.
Antiferromagnetic phases correspond to a checkerboard of $+1$'s and $-1$'s in the plane, which
is ordered behavior of agents.  The agents with potential \eqref{E:mgLongShort} have no such
ordered behavior: $+1$'s and $-1$'s are seen equally but in a random unordered fashion,
which reflects that supply is equally accessible to everyone.

There are four types of phases in the dipolar model 
(cf. \cite{CGT} for details and references).  
The first two occur below a critical temperature $T_c(\Delta)$ that depends on 
$\Delta=\delta/b$.
When $\Delta$ is smaller than $\Delta_a\approx 0.425$, the ground state is the
antiferromagnetic one,
consisting of a checkerboard of $+1$ and $-1$ values of the $\sigma$.  When $\Delta>\Delta_a$,
the transition is from the antiferromagnetic state to a state consisting of
antialigned stripes of width one, representing large areas of buyers and large areas of
sellers that persist in some direction.  As $\Delta$ gets larger, the stripes grow in width.
For example, stripes of widths two, three, and four appear at $\Delta_k\approx 1.26,2.2,2.8$,
respectively. 

The other two phases occur when $T\geq T_c(\Delta)$. 
There is a transition to a disordered \emph{tetragonal} phase above and near the
$T_c(\Delta)$, and this phase is not paramagnetic.   
There are extended ferromagnetic domains characterized by
predominantly square corners, and a typical configuration looks like a maze of $+1$ trails
and $-1$ walls.  For larger temperatures, the transition is to a paramagnetic phase.
Similar phase behavior occurs for the $w=1$ Coloumb model in \eqref{E:mgDipoleShort} \cite{MM}.

The antiferromagnetic phase is one with buyers and sellers in a checkerboard
pattern.  This happens when the demand $p(Q)$ is mostly due to nearest-neighbor oligopolistic
competition; i.e., the terms $-b\sum_{i:|i-k|=1} q_k q_i$ dominate agent $k$'s demand function.
When there are stripes there are large areas persisting in some direction that are buying 
and other such areas that are selling.  In this case, agent $k$ has localized oligopolistic
competition in one direction and heightened local competition in the other direction.
As the stripes widen, there is heightened local competition within the stripes, and a localized
oligopolistic competition among the groups of agents in the stripes.
In the tetragonal phase, there are large unordered areas of buyers and sellers that look
like a maze.   In the paramagnetic phase, buyers and sellers tend to appear at random, and this
corresponds to the standard Cournot model of nonlocalized oligopolistic competition.

It seems intuitive that \eqref{E:mgLongShort} would not display stripes, but rather be fully
frustrated or completely aligned; that is, if supply is globally accessible and local prices
are essentially the same for all agents (such as a homogeneous market with no trade barriers
and uniform distribution of goods)
then there should not be large clumps of buyers and sellers.  On the other hand, if there are
no local interactions as in \eqref{E:mgPotential}, then there should only be antiferromagnetic
states or fully frustrated states with high probability, and no large clumps of buyers or
sellers.

The numerical evidence in
\cite{LVS, PSBS} for the inductive minority game \eqref{E:inductiveMG} suggests it is
insensitive to the types of long range interactions; specifically that the types in
\eqref{E:mgPotential} of equal interaction strength show the same behavior as the falloff
types in \eqref{E:mgDipoleShort} with $\delta=0$.  This is in contrast to the deductive model
presented here. 

The stochastic nature of public information in the inductive model does induce a phase
transition.  It would be interesting to determine if the deductive model presented here with
a random field component, which could be represented by random marginal cost fluctuations
(cf. \secRef{S:Cournot}), will yield a phase
transition.  Such random costs could, for example, represent the fees, which vary from agent
to agent, charged to each agent per buy/sell.  In \cite{PGGS}, the addition of demand into a
model of stock trade is analyzed by introducing an effective field as mentioned above.

Evidence is found in \cite{PSBS} that the dynamics of the inductive minority game is intrinsic
to resource allocation.  The view of the minority game as a potential game 
makes clear why this is the case, in
the sense that all potential games are isomorphic to congestion problems \cite{MS}, and the
potential for the minority game is a specific, discrete case of a supply-and-demand potential
(c.f. \secRef{S:Cournot}).

%
%
%
%
%
\section{Conclusion} \label{S:Conclusion}
We have seen that potential games (with nice potentials), when coupled with deviations-from-rationality,
result in models for which equilibrium values of all  relevant (macroscopic) quantities
such as output per agent, payoff per agent, etc., are determined by averaging over a Gibbs
measure with the given potential at inverse temperature $\beta$.  The introduction of 
temperature in economics simply corresponds to the assumption that players deviate from
the purely rational behavior of producing output in the direction of maximum increasing
payoff.

A very nice feature of the Gibbs approach is that it is accessible and demonstrates different
patterns of behavior among agents for different types of distribution of goods.  This is
desirable when modeling trade barriers, shipping costs, and real-life models such as
restaurants, where demand is localized.  It can also be used in conjunction with renormalization
techniques, which could model large-scale behavior of `agglomerated' agents, analogous to
block spins for physical systems.

It would also be of interest to pursue more depth insofar as the economic implications of
different types of phases for the different models presented here.  Since the author is
not an economist, such an endeavor will be left to the experts.

\ack{I wish to thank Alexander Chorin for exposing me to fascinating ideas, which ultimately
led to this paper.}

%
%
%
%
%
%
\appendix

\newpage
%
%
%
%
\section{Proof of Vector Fokker-Planck Equation} \label{S:appendixA}
	Here we will show the derivation of a vector Fokker-Planck equation for the joint distribution given the Langevin dynamics in \cite{AGH1}.  This proof is just a slight modification of the proof in \cite{AGH1}.  We only consider a finite number of agents $\np<\infty$.

In a potential game, the It\^o equation for dynamics is $d\vx = \grad{V}dt + \sigma d\vw(t)$ as in \eqref{E:Langevin}.  For a small change in time $\Delta t$, the dynamics can be written
	\begin{equation}
	\Delta\vx(t) \equiv \vx(t+\Delta t) - \vx(t)
		= \grad{V}\Delta t + \sigma\Delta\vw(t) + \vec{o}(\Delta t),
	\label{E:langevinFinite}
	\end{equation}
with $\sigma\Delta\vw(t)$ a normal random variable of mean zero and variance $\sigma^2\Delta t$, and $\vec{o}(\Delta t)$ is the usual vector `little order' notation.
	\footnote{
		    $\vec{v}(\Delta t)$ is $\vec{o}(\Delta t)$ if 
		    $\lim_{\Delta t\to 0} v_i(\Delta t)/(\Delta t) = 0$ for each component 
		    $v_i$ of $\vec{v}$.}
The configuration of decisions $\vx(t)$ at time $t$ is hence a random variable with time-dependent joint density $f(\vx,t)$.  Let $h(\vx):\Omega\to(-\infty,\infty)$ be an arbitrary twice differentiable function such that $h$ and $\grad h$ vanish on the boundary $\partial\Omega$ of decision space $\Omega$.  Note $h$ does not explicitly depend on time.  Then the expected value of $h(\vx)$ (over phase space) at time $t+\Delta t$ is
	\begin{equation}
	E\left[ h( \vx(t+\Delta t) )\right] = \int_\Omega h(\vx)f(\vx,t+\Delta t) d\vx,
	\label{E:expectedh}
	\end{equation}
where $d\vx$ is the product measure $dx_1dx_2\cdots dx_\np$ on $\Omega$.
Note above that $\vx(t+\Delta t)$ on the left side represents the random variable over which the expected value $E$ is to be taken
	\footnote{
		    Rigorously, the map $\vx(t)\to \vx(t+\Delta t)$ generates a differentiable global
		    (stochastic) flow on decision space $\Omega$ for nice potentials, and the global flow has
		    a generator which will be shown below.}, 
whereas the $\vx$'s on the right side are the points in $\Omega$.  Equation \eqref{E:langevinFinite} can be used to obtain another expression for \eqref{E:expectedh}
	\footnote{
		    For simplicity, we use the formal notation of differential  stochastic calculus.}:
	\begin{equation}
	E\left[ h( \vx(t+\Delta t) )\right] = E\left[ h( \vx(t)+\Delta x(t) ) \right]
	= E\left[ h(  \vx(t)+ \grad{V}(\vx(t))\Delta t + \sigma\Delta\vw(t) ) \right]
	+\vec{o}(\Delta t).
	\label{E:expectedh2}
	\end{equation}
The expression on the right hand side of \eqref{E:expectedh} will be subtracted from the Taylor expansion of the right side of \eqref{E:expectedh2} to yield the Fokker-Planck equation.  We will proceed as follows.

Let $g(\vy)$ be the (joint) density of $\sigma\Delta\vw(t)$, which is an $\np$-dimensional normal density with mean zero and variance $\sigma^2\Delta t$.  The left side of \eqref{E:expectedh2} is in terms of the random variables $\vx(t+\Delta t)$, and the right side is in terms of the variables $\vx(t)$ and $\vy$ with respective densities $f(\vx,t)$ and $g(\vy)$.  Accordingly, the right side of \eqref{E:expectedh2} can be written as an integral over these densities:
	\begin{equation}
	E\left[ h( \vx(t+\Delta t) )\right] = 
	\int_{\mathbb{R}^n}\int_\Omega
	  h\left( \vx(t) + \grad{V}(\vx(t))\Delta t + \sigma\vy \right)
	f(\vx,t) g(\vy) d\vx d\vy,
	\label{E:expectedh3}
	\end{equation}
where $\mathbb{R}=(-\infty,\infty)$.

The Taylor expansion of the right side of \eqref{E:expectedh3} yields
	\begin{equation}
	\begin{aligned}
	\int_{\mathbb{R}^n} \int_\Omega
	&\Big\{
	 h(\vx(t)) + \grad{h}(\vx(t))\cdot[\grad{V}(\vx(t))\Delta t + \vy]
	\\
	 &\:+ \frac{1}{2}[\grad{V}(\vx(t))\Delta t + \vy]\cdot
		D^2h(\vx(t)) \cdot[\grad{V}(\vx(t))\Delta t + \vy]^T + \vec{o}(\Delta t)
	\Big\}
	f(\vx,t)g(\vw)d\vx \, d\vy,
	\label{E:TaylorE}
	\end{aligned}
	\end{equation}
where $D^2h$ is the second-derivative matrix for $h$ (i.e., the $ij$ entry is $\partial^2h/\partial x_i\partial x_j$) and the transpose of a row vector $\vec{v}$ is the column vector $\vec{v}\,^T$. 

Integrating \eqref{E:TaylorE} over the $\vy$ eliminates terms linear in $y_i$ as well as the mixed second-order terms since the $y_i$ have mean zero and are independent.  The expected value of $w_i^2$ is $\sigma^2\Delta t$, hence \eqref{E:TaylorE} reduces to
	\begin{equation}
	\begin{aligned}
	\int_\Omega h(\vx(t)) f(\vx,t)dx 
	&+ \Delta t\int_\Omega \grad{h}(\vx(t))\cdot\grad{V}(\vx(t)) f(\vx,t) d\vx
	\\
	&+ \Delta t \frac{\sigma^2}{2} \int_\Omega \nabla^2 h(\vx(t))  f(\vx,t) d\vx
	+ o(\Delta t),
	\end{aligned}
	\label{E:TaylorEInty}
	\end{equation}
where $o(\Delta t)$ is the one-component version of $\vec{o}(\Delta t)$ and $\nabla^2$ is the Laplacian.  Thus the generator for the Langevin dynamics $\vx(t+t_0)=\mathfrak{L}^t \vx(t_0)$ is $\mathbb{L}=(\grad V)\cdot\grad + (\sigma^2/2) \nabla^2$, where $\mathbb{L}$ acts on twice-differentiable functions on $\Omega$ that vanish on the boundary.  It is easy to show $\mathbb{L}$ is symmetric and the domain of its adjoint is the Sobolev space $H^2(\Omega)$ for a well-behaved potential $V$.  We view its adjoint $\adj{\mathbb{L}}$ as a densely defined unbounded operator in $L^2(\Omega,d\vx)$.  The idea is apparent from the following.

Using Green's identities, the integrals containing the terms $\grad h$ and $\nabla^2 h$ can be integrated by parts, and the boundary integrals vanish on the boundary $\partial\Omega$ (since $h$ and $\grad h$ do) leaving a new expression for \eqref{E:TaylorEInty}:
	\begin{equation}
	\begin{aligned}
	\int_\Omega h(\vx(t)) f(\vx,t)dx 
	&- \Delta t\int_\Omega h(\vx(t)) \grad\cdot[\grad{V}(\vx(t)) f(\vx,t)] d\vx
	\\
	&+ \Delta t \frac{\sigma^2}{2} \int_\Omega h(\vx(t)) \nabla^2 f(\vx,t) d\vx
	+ o(\Delta t).
	\end{aligned}
	\label{E:TaylorEInty2}
	\end{equation}
This shows that the adjoint of $\mathbb{L}$ operates as $\mathbb{L}^*(f)=-\grad\cdot(f \grad V) + (\sigma^2/2) \nabla^2 f$.  Equations \eqref{E:expectedh} and \eqref{E:TaylorEInty2} are identical, hence subtracting them yields
	\begin{equation}
	\begin{aligned}
	\int_\Omega h(\vx(t)) [f(\vx,t+\Delta t) - f(\vx,t)]dx =
	\Delta t\int_\Omega h(\vx(t)) [-\grad\cdot[\grad{V}(\vx(t)) f(\vx,t)]
		+ \frac{\sigma^2}{2} \nabla^2 f(\vx,t) ] d\vx
	+ o(\Delta t).
	\end{aligned}
	\label{E:FPInt}
	\end{equation}
Dividing \eqref{E:FPInt} by $\Delta t$, taking the limit $\Delta t\to 0$ results in the Fokker-Planck equation \eqref{E:FokkerPlanck} since the set of all such $h$'s described are a separating set for probability densities on $\Omega$ (i.e., these $h$'s can weakly approach delta functions on $\Omega$).

Let $h(\vx)$ be an observable (i.e. function) on $\Omega$ that is differentiable, vanishes on $\partial\Omega$, and doesn't depend explicitly on time.  For a stationary state $f$, the average of $h(\vx)$ is constant in time:
 	\begin{equation}
	\begin{aligned}
	0 &= \frac{\partial}{\partial t} \int_\Omega h(\vx) f(\vx,t) d\vx
	\\
	  &= \int_\Omega h(\vx) \grad\cdot\left\{-[\grad{V}(\vx(t)) f(\vx,t)] 
				+ \frac{\sigma^2}{2} \grad f(\vx,t)\right\} 
	\\
	  &= -\int_{\Omega} \grad h(\vx) \cdot \left\{-[\grad{V}(\vx(t)) f(\vx,t)] 
				+ \frac{\sigma^2}{2} \grad f(\vx,t)\right\}. 
	\end{aligned}
	\label{E:FokkerPlanck2}
	\end{equation}
Since $h$ was arbitrary and $\grad h$ can be made to approach delta functions, for a stationary state
	\begin{equation}
	-[\grad{V}(\vx(t)) f(\vx,t)] + \frac{\sigma^2}{2} \grad f(\vx,t) = 0.
	\label{E:statState}
	\end{equation}
We then see that the Gibbs state in \eqref{E:Gibbs}	
is the unique solution to \eqref{E:statState}, and is the equilibrium state for the potential game.
The negative of the Helmholtz free energy is in fact a Liapunov function for the dynamics and it can be used to show (see \cite{AGH1}) that the unique finite-agent solution \eqRef{E:Gibbs} of \eqRef{E:statState} is attained in the long run.

\newpage

%
%
%
%
%
%
\section{Solution of the Cournot Free Energy} \label{S:appendixB}
The free energy for the Cournot oligopoly and collusion is obtained from the partition function in \eqRef{E:partitionCO}
	\begin{equation}
	\mathcal{Z}_{\np}\!\!\left(\tilde{b}\right) = \int \e^{\beta V}
	= \int_{\qmin}^{\qmax} 
	  \exp\left[ -\beta\frac{b}{\np} \sum_{j,k=1}^{\np} \tilde{q}_j \tilde{q}_k
		-\beta\frac{\tilde{b}}{\np}\sum_{j=1}^{\np} \tilde{q}_j^2 
		+ \beta(a-c)\sum_{j=1}^{\np} \tilde{q}_j \right]
	  \prod_{j=1}^{\np} d\tilde{q}_j.	
	\label{E:partitionCO2}
	\end{equation}
The limit of the free energy 
	\begin{equation}
      F_{\np} \left(\tilde{b}\right)=\frac{1}{\beta\np}
		\ln\!\left(\mathcal{Z}_{\np}\!\!\left(\tilde{b}\right)\right)
	\label{E:freeEnergy}
	\end{equation}	
will be shown below.  First note that the limit does not depend on the value of $\tilde{b}$, since
	\begin{equation}
	-\frac{\tilde{b}\qmax^2}{n} + F_{\np} 
	\leq 
	F_{\np} \left(\tilde{b}\right)
	\leq 
	-\frac{\tilde{b}\qmin^2}{n} + F_{\np},
	\end{equation}
where $F_{\np}\equiv F_{\np}(0)$.  It then suffices to set $\tilde{b}$ to zero in \eqRef{E:partitionCO2}, which will be done below to simplify a limit.
For further manipulation, changing variables to $q=\tilde{q}-\gamma$, $\gamma\equiv(\qmax+\qmin)/2$, $h\equiv a-c$, results in
	\begin{equation}
	\begin{aligned}
	\mathcal{Z}_{\np} 
	= &\exp\left[ -\beta b\gamma^2 n - \beta\tilde{b}\gamma^2+ \beta h\gamma n 
			  + \beta \frac{n}{b} \left( 
			       \gamma b + \gamma \frac{\tilde{b}}{\np} - \frac{h}{2} 
			    \right)^2
	     \right]
	\\
	&\;\times  
	\int_{[-Q/2,Q/2]^\np} \exp\left[
	  -\beta \left( \sqrt{\frac{b}{\np}} \sum q_j 
	  + \sqrt{\frac{\np}{b}} \left( b\gamma + \frac{\tilde{b}}{\np}\gamma - \frac{h}{2} \right)
		\right)^2
	  -\beta\frac{\tilde{b}}{\np}\sum q_j^2 \right]
	  \prod_{j=1}^{\np} dq_j,
	\end{aligned}
	\label{E:partitionCO3}
	\end{equation}
where $Q\equiv\qmax-\qmin$.

The solution can be found using a saddle point method, and the key is to use the identity
	\begin{equation}
	\e^{-\beta p^2} = \frac{1}{ \sqrt{\pi}\sqrt{4\beta} }
	\int_{-\infty}^{\infty}
	   \exp\left( -\frac{1}{4\beta} t^2 + ipt \right) dt
	\label{E:approx}
	\end{equation}
as with the simplest case in \cite{K}.
Using the above in \eqref{E:partitionCO3} and setting $\eta\equiv b\gamma + \tilde{b}\gamma/n - h/2$ results in
	\begin{equation}
	\begin{aligned}
	\mathcal{Z}_\np 
	= \frac{1}{ \sqrt{\pi} \sqrt{4\beta} }
		&\exp\left[ \frac{\beta h^2 \np}{4b} + \beta \tilde{b} \gamma \left( 
			\frac{ \tilde{b}\gamma }{ b \np} + \gamma  - \frac{h}{b} \right)
	     \right]
	\int_{-\infty}^{\infty} 
		\exp\left[\frac{-1}{4\beta}t^2 
			+ i\sqrt{\frac{\np}{b}} \eta t
		\right]
	\\
	&\;\times  
		\left\{ \int_{-Q/2}^{Q/2}  
	  \exp\left[-\beta \frac{\tilde{b}}{\np} q^2 \right]
	  \exp\left[ i \sqrt{\frac{b}{\np}}q t \right]
	  dq
	\right\}^\np dt,
	\end{aligned}
	\label{E:par2}
	\end{equation} 	
where the order of integration was changed via Fubini's theorem.  Consider part of the integrand above
	\begin{equation}
	g_{\np}\!\left( \sqrt{\frac{b}{\np}}\,t \right) 
	= 
	\frac{1}{Q} \int_{-Q/2}^{Q/2}
		\exp\left[-\beta \frac{\tilde{b}}{\np} q^2 \right]
	  	\exp\left[ i q \sqrt{\frac{b}{\np}}t \right]
	dq
	= \sinc\!\!\left( \frac{Q}{2}\sqrt{\frac{b}{\np}}\,t \right),
	\label{E:g}
	\end{equation}
where $\sinc(x)\equiv\sin(x)/x$ ($\sinc(0)\equiv 1$), $\tilde{b}$ was explicitly set to zero, and a normalizing factor was added so that $g_t(0)=1$.
Changing variables to $\hat{t}=t/\sqrt{n}$, the partition function can then be written
	\begin{equation}
	\begin{aligned}
	\mathcal{Z}_{\np} 	 
	= 
	&\exp\left[-\beta\gamma(b\gamma - h)\np\right]
	\frac{ Q^{\np} \sqrt{\np} }{ \sqrt{\pi}\sqrt{4\beta} }
	\\
	\times
	  &\;\int_{-\infty}^{\infty}
		\left\{
		\exp \left[ -\frac{1}{4\beta} 
		 \left( \hat{t} - i\frac{2\beta}{\sqrt{b}} \eta \right)^2
		\right]
		\sinc\!\!\left( \frac{Q}{2}\sqrt{b}\,\hat{t} \right) 
		\right\}^{\np}
		\,d\hat{t}.
	\end{aligned}
	\label{E:par3}
	\end{equation}
Now a basic fact is needed to determine the limit of the free energy.

%
%
%
%
\begin{lemma}\label{L:converge}
Let $f(x)$ be a bounded, continuous, real-valued, integrable function on the real line 
$\mathbb{R}$ and let $G$ be an open set.  Suppose there is a point $x_0\in G$ for which
$|f(x)|<f(x_0)$ for all $x\not= x_0$ and $\limsup|f(x)|<f(x_0)$ for all $x$ outside of
some neighborhood of $x_0$ contained in $G$.  Then
	\begin{equation}
	\lim_{m\to\infty} 
	  \left\{  \int_G \left[ f(x) \right]^{m+k} \,dx  
	  \right\}^{1/m}
	= f(x_0)
	\label{E:fnLim}
	\end{equation}  
for any fixed integer $k$.
Furthermore, for a function $g(x)$ that is continuous at $x_0$ with $gf^k\in L^\infty$,
	\begin{equation}
	\lim_{m\to\infty}
	\frac{\int_{-\infty}^{\infty} g(x) \left[ f(x) \right]^m \,dx
	    }{\int_{-\infty}^{\infty} \left[ f(x) \right]^m \,dx}
	= g(x_0).
	\label{E:measureLim}
	\end{equation}
\end{lemma}
\begin{proof}  
First, by factoring $\|f\|_{\infty}$ out of the integral in \eqref{E:fnLim}, we may
assume $f(x_0)=1$
by considering the function $f/\|f\|_{\infty}$.  Choose a number $\epsilon_0>0$ so that 
$\limsup|f(x)|<1-2\epsilon_0$ for all $x$ outside of a neighborhood 
$\mathcal{N}\subset G$ of $x_0$ and $f>0$ on $\mathcal{N}$.
Let $S_{1-\epsilon_0}=\left\{ x\in G | f(x)\geq 1-\epsilon_0 \right\}$, and
$I_m=( \int_G [ f(x) ]^{m+k} \,dx )^{1/m}$.
Since $I_m \leq f(x_0)^{(m+k-2)/m} ( \int_G f^2(x)\,dx )^{1/m}$ for large $m$,
$\limsup_{m\to\infty} I_m \leq f(x_0)$.  The inequality
	\begin{equation}
	(I_m)^m \geq (1-\epsilon)^{m+k-2} \int_{S_{1-\epsilon}} f^2(x)\,dx
		- (1-2\epsilon_0)^{m+k-2} \int_{G-\mathcal{N}} f^2(x)\, dx
	\end{equation}
holds for all $\epsilon\leq\epsilon_0$.  Upon factoring out $(1-\epsilon)^{m+k-2}$, 
it is seen that $\liminf_{m\to\infty} I_m\geq f(x_0)$.
To prove \eqref{E:measureLim}, the following is needed: 
	\begin{equation}
	\lim_{m\to\infty}
	\left|\frac{\int_{\mathbb{R}-S_{1-\epsilon}} g(x) \left[ f(x) \right]^m \,dx
	          }{\int_G \left[ f(x) \right]^m \,dx}
	\right|
	\leq
	\|gf^k\|_\infty  \lim_{m\to\infty}
	\frac{\int_{\mathbb{R}-S_{1-\epsilon}} |f(x)|^{m-k} \,dx
	    }{\left|\int_G \left[ f(x) \right]^m \,dx\right|}
	=0,
	\label{E:meas1}
	\end{equation}
where the last equality follows from the root test for convergence since the
hypothesis implies the limit of the ratio on the right side of \eqref{E:meas1}
to the power $1/m$ is less than one.  With this, \eqref{E:measureLim} reduces to 
	\begin{equation}
	\frac{\int_{S_{1-\epsilon}} g(x)[f(x)]^m\,dx
	    }{\int_{\mathbb{R}} [f(x)]^m\,dx}.
	\end{equation}
By the continuity of $g$ and \eqref{E:meas1} again with $G=S_{1-\epsilon}$, it is seen
that \eqref{E:measureLim} holds.
\end{proof}

To apply \lemRef{L:converge} to the partition function in \eqref{E:par3}, the integration
path along the real line must be deformed in the complex plane to a path on which the 
integrand is real-valued.  As such, consider the function 
$f(z)=\exp\left[-k(z +i \eta_0)^2\right]\sinc(z)$, where $\eta_0= -\beta Q\eta$ and
$1/k=\beta bQ^2$ (changing variables $t=Q\sqrt{b}\,\hat{t}/2$ in \eqref{E:par3}; note 
$\eta_0$ is increasing in $h$).

Let $z=x+iy$, and $f(z)=\mathcal{R}(x,y)+i\mathcal{I}(x,y)$, where 
$x,y,\mathcal{R},\mathcal{I}$ are real-valued.  Then we are interested in the paths 
$(x(s),y(s))$ in the complex plane $\mathbb{C}$ on which $\mathcal{I}=0$.  Since $f$ is
holomorphic and not identically zero, the gradient of $\mathcal{I}$, $\grad\mathcal{I}$,
can only vanish on a closed set of isolated points in $\mathbb{C}$.  Thus $\mathcal{I}^{-1}(0)$
consists of various differentiable paths in $\mathbb{C}$.  The imaginary part can be
expanded
	\begin{equation}
	\begin{aligned}
	\mathcal{I}(x,y) =
		\frac{ \e^{-kx^2 + k(y+\eta_0)^2} }{ x^2+y^2 }
		&\big\{ \sin(-2kx(y+\eta_0))[x\sin x \cosh y + y\cos x \sinh y]
		\\
			&+ \cos(-2kx(y+\eta_0))[-y\sin x \cosh y + x\cos x \sinh y]
		\big\}.
	\end{aligned}
	\label{E:I}
	\end{equation}

%
%
%
%
\begin{lemma}\label{L:deform}
  There is a piecewise differentiable path $z(s)$ such that $\mathcal{I}(z(s))=0$.
In addition, $z(s)$ can be written $(x,y_z(x))$ for small and large $|x|$, with 
$\lim_{x\to\pm\infty}y_z(x)=C_1$ and $\lim_{x\to 0}y_z(x)=C_2$ for constants
$C_1$ and $C_2$.  
\end{lemma}
\begin{proof}
Consider fundamental domains in the $z$-plane of $w=\sinc(z)$.
The fundamental domains $D_j$, $D'_j$ shown in figure \ref{F:lnsinc} are then
separated by curves satisfying $y=0$ and $\tan x/x=\tanh y/y$, with $x\not=0$.  
If $x>0$, the implicit function theorem implies the latter
curves can be written $x=c_j(y)$ with $c_j$
differentiable and $j\pi<c_j(y)<(2j+1)\pi/2$, $j=1,2,\dots$.
Let $c_0(y)=0$ and for $x<0$, the curves are reflected: $c_{-j}=-c_j$.
The branch cuts in the $w$-planes are then the images of $c_{2j+1}$.
The logarithm $\ln(\sinc z)$ will now map two interior points in the $z$ plane
to a complex number since $\sinc(-z)=\sinc(z)$.
Branch cuts at $|x|>0$ are added in the $z$-plane on the $x$-axis to 
avoid discontinuities in $\alpha$.
The fundamental domains $D_j$ lie below the $x$-axis between the curves $c_{2j-1}$
and $c_{2j+1}$, whereas the $D'_j$ lie above the $x$-axis between $c_{-2j-1}$ 
and $c_{-2j+1}$.
The shaded regions of the $D_j$ and $D'_j$ in the figure below
correspond to the lower half of the $w$-planes.

Let $\alpha=\alpha(x,y)$ be the angle of $\sinc(z)$ (i.e., the imaginary
part of $\ln(\sinc z)$).  Figure \ref{F:lnsinc} shows that in the lower half
of the $z$-plane, $\alpha$ increases in $x$ and is of order
$x$.


\begin{figure}[h]
\begin{center}
\includegraphics[width=7in,height=5in]{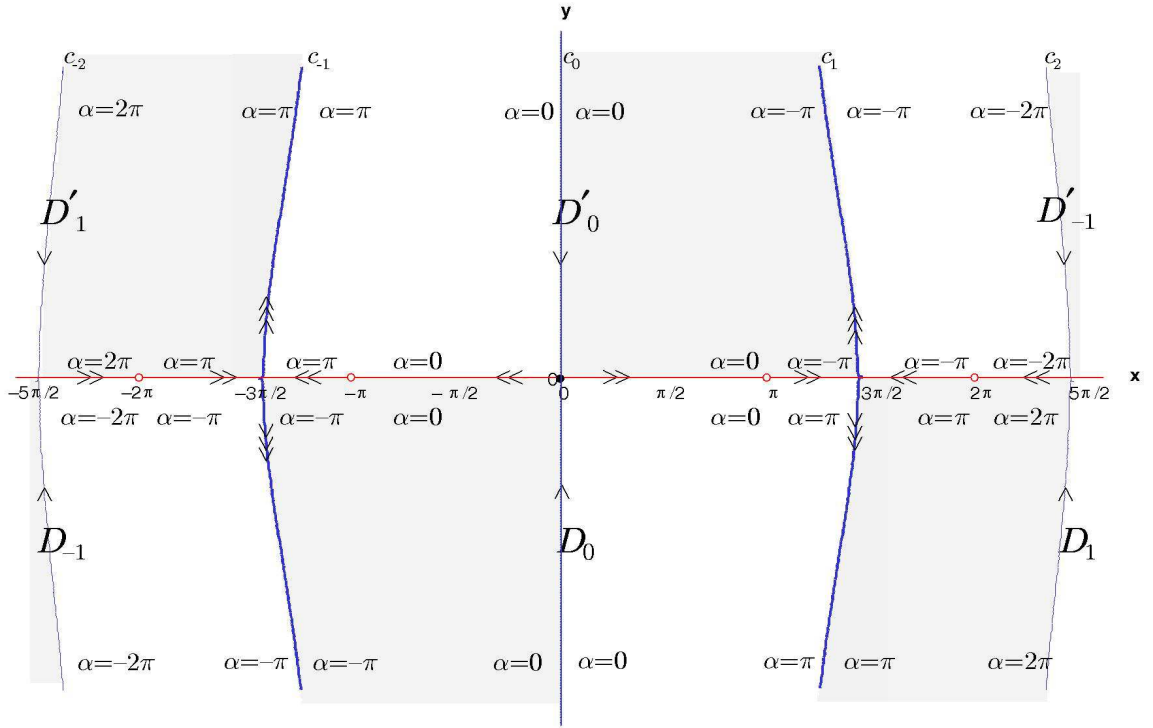}
\caption{$\ln(\sinc z)$ values}
\label{F:lnsinc}
\end{center}
\end{figure}

Differentiating the identity for $\tan\alpha$,
	\begin{align}
	\frac{\partial\alpha}{\partial x} &=
	\frac{2y\left( \sin^2x+ \sinh^2y \right) -\left(x^2+y^2\right)\sinh2y}
		{2\left(x^2+y^2\right)\left(\sin^2x+\sinh^2y\right)}
	\label{E:alphax}
	\\
	\frac{\partial\alpha}{\partial y} &=
	\frac{-2x\left( \sin^2x+ \sinh^2y \right) +\left(x^2+y^2\right)\sin2x}
		{2\left(x^2+y^2\right)\left(\sin^2x+\sinh^2y\right)}
	\label{E:alphay}
	\end{align}
and it is seen that $\frac{\partial\alpha}{\partial x}>0$ ($<0$) when $y<0$ ($>0$). 

The part in braces in \eqref{E:I} can be written 
	\begin{equation}
	\sqrt{x^2+y^2}\sqrt{\sin^2 x + \sinh^2 y}
	\,\sin(-2kx(y+\eta_0)+\alpha),
	\label{E:Ireduced}
	\end{equation}
and this is zero if and only if $\mathcal{I}$ is zero.  For the case $\eta_0=0$, the integral
in \eqref{E:par3} can be evaluated using \lemRef{L:converge}, with the maximum of the integrand
occuring at zero. In the case $\eta_0\not= 0$,  the paths in $\mathbb{C}$ on which $f(z)$ is 
real-valued is precisely where 
	\begin{equation}
	-2kx(y+\eta_0)+\alpha = m\pi,
	\label{E:realF}
	\end{equation}
where $m$ is an arbitrary integer.

Since $y+\eta_0 = \alpha/(2kx) - m\pi/(2kx)$ and $\alpha$ is of order $x$, it is evident
that $|y|\to\infty$ as $x\to 0$ on such paths unless $m=0$.  This is not surprising, since
large $|z|$ behavior of $f$ is dominated by the $z^2$ term in the exponential (thus giving
hyperbolae-type curves).  We will single out the $m=0$ curve $z(s)=(x(s),y(s))$ satisfying 
	\begin{equation}
	y +\eta_0=\alpha(x,y)/(2kx)
	\label{E:newCurve}
	\end{equation}
for reevaluating the integral in \eqref{E:par3}.
Since $\alpha$ is of order $x$, the $y$-component of $z(s)$, $y(s)$, is bounded and has
finite limits as $x\to\pm\infty$.  The location of the curve depends on $\eta_0$.
To see this, note that $\alpha(x,y)<Dx$ for a constant $D$ with
$|D|\leq 1$, and the sign of $D$ is negative in the upper-half and positive in the
lower-half of the $z$-plane.  If $\eta_0>0$ and the curve $z(s)$ were in the upper-half
of the $z$-plane, then $y=\alpha/(2kx)-\eta_0\leq D/(2k) - \eta_0 <0$, which is
a contradiction.  Hence if $\eta_0>0$ ($<0$), then the curve $z(s)$ is in the lower (upper)
half of the $z$-plane. 

The $y$-intercept of $z(s)$ can be found using $\lim_{x\to 0^+}\alpha=0$ 
and $\lim_{x\to 0} \sin(\alpha)/\alpha =1$.  Near $x=0$, $z(s)$ can be written
$z=(x,y_z(x))$, from which
	\begin{equation}
	\begin{aligned}
	y_z\!(0)+\eta_0 = \lim_{x\to 0} \frac{\alpha}{2kx}
	&=
	\lim_{x\to 0} \frac{-y_z \sin x \cosh y_z + x\cos x \sinh y_z}
		{ 2kx\sqrt{x^2+y_z^2}\sqrt{\sin^2 x \cosh^2 y_z + \cos^2 x \sinh^2 y_z} }
	\\
	&= \frac{-y_z\!(0)\cosh y_z\!(0) + \sinh y_z\!(0)}{2ky_z\!(0)\sinh y_z\!(0)}.
	\end{aligned}
	\label{E:curvey}
	\end{equation}
The solutions of the above equation in $y_z\!(0)$ are the same as the solutions in $y$ to the equation $y/[1-2k(y+\eta_0)y]=\tanh(y)$, except for an extraneous root of $y=0$ in the latter (since $\eta_0\not= 0$).  The solution for the $y$-intercept is unique, and this will be shown
later.
\end{proof}

With the above, the integral in \eqref{E:par3} can be deformed onto the curve $z(s)$.
To apply \lemRef{L:converge},
it is necessary to find the maximum of $|f|$ on the full curve where $\mathcal{I}=0$, and to
also show it occurs at a single point $z_m$ such that $f(z_m)>0$.

Equivalently, it will be easier to find the maximum of 
	\begin{equation}
	|f|^2=\exp\left[-2kx^2+2k(y+\eta_0)^2\right]
		\left(\sin^2x\cosh^2y+\cos^2x\sinh^2y\right)/(x^2+y^2)
	\end{equation}
on the full $\mathcal{I}=0$ curve $z(s)=(x(s),y(s))$.  
Since $\lim_{s\to\pm\infty}f(z(s))=\lim_{x\to\pm\infty}f(x,y_z(x))=0$, a maximum occurs and is greater than zero.  On the curve near this maximum, say at $s_m$, the argument of $f$ is constant (either $\Arg(f)=0$ or $\Arg(f)=\pi$) hence $\tan(\Arg(f))=\mathcal{I}/\mathcal{R}$ is constant.  Since $\tan(\Arg[f(x(s),y(s))])'=0$, the tangent vector $\vec{v}(s)$ to the curve $(x(s),y(s))$ is orthogonal to $\grad(\mathcal{I}/\mathcal{R})$ near $s_m$.

Any maximum point of $|f|^2$ on the curve $z(s)$ must satisfy
	\begin{equation}
	\begin{aligned}
	0=\frac{\partial}{\partial x}|f|^2 = 
			&\frac{\exp[-2x^2+2(y+\eta_0)^2]}{(x^2+y^2)^2}
		\\
		  &\times
		   \left\{ -2x(2kx^2+2ky^2+1)\left(\sin^2x\cosh^2y+\cos^2x\sinh^2y\right)
				+ (x^2+y^2)\sin2x
		   \right\}
	\end{aligned}
	\label{E:x}
	\end{equation}
and
	\begin{equation}
	\begin{aligned}
		0=\frac{\partial}{\partial y}|f|^2 = 
			&\frac{\exp[-2x^2+2(y+\eta_0)^2]}{(x^2+y^2)^2}
		\\
		  &\times
		   \left\{(4k(y+\eta_0)(x^2+y^2)-2y)
			\left(\sin^2x\cosh^2y+\cos^2x\sinh^2y\right)
			+ (x^2+y^2)\sinh2y
		   \right\}.
	\end{aligned}
	\label{E:y}
	\end{equation}

%
%
\begin{lemma}\label{L:saddleRegion}
The conditions in \eqref{E:x} and \eqref{E:y} imply that the maximum of $|f|^2$
on the curve $z(s)$ occurs at $(0,y_z(0))$, the $y$-intercept of the curve $z(s)$. 
\end{lemma}
\begin{proof}
Without loss of generality it can be assumed that $\eta_0\geq0$, since the integral in \eqref{E:par3} is an even function of $\eta_0$.  

For the case $\eta_0=0$, the curve $z(s)$ is the $x$-axis and 
$|f|^2(z(s))=\e^{-2kx^2}\sinc^2x$.  Clearly $(0,0)$ is a global maximum of $|f|^2$
on $z(s)$.  Hereon, the case $\eta_0>0$ is considered.

We will consider the case $x\not= 0$, and $x=0$ will be dealt with subsequently.  Then \eqref{E:x} is equivalent to
	\begin{equation}
	1 > \sinc2x
	  = \frac{2kx^2 + 2ky^2 + 1}{x^2+y^2}
		(\sin^2x + \sinh^2y)
	  \geq 0.
	\label{E:xCond1}
	\end{equation}

The inequality $\sinc2x\geq 0$
implies $x\in[-\pi/2,\pi/2]$ or $x\in\pm[m\pi,(m+1)\pi/2]$ for positive integers $m$.
In order for \eqref{E:xCond1} to hold, $\sin^2x+\sinh^2y \leq \sinc(2x)(x^2+y^2)$, which
is impossible for $|x|\leq\pi/2$.  Any extreme points of $|f|^2$ for which $x\not= 0$
must therefore satisfy $|x|\geq\pi$.

From the above argument, maximum points of $|f|^2$ in the set 
$S_0=\{(x,y):|x|\leq\pi/2\}$ must satisfy $x=0$.  
Any maximum point $(x,y)\notin S_0$ will then satisfy $|x|\geq\pi$.  
It is seen from \eqref{E:newCurve} that $y+\eta_0<0$ when $x>0$ and $y<0$, which
along with \eqref{E:y}, implies that $|y|>(\pi^2+y^2)\tanh|y|$ at an extreme point. 
That this is impossible shows the largest value of $|f|^2$ on the curve $z(s)$
will occur at $x=0$.

It only remains to show the intersection of $z(s)$ with the $y$-axis is unique and
is also a maximum point of $|f|^2$.  Note that \eqref{E:x} is satisfied when $x=0$.
Also note that \eqref{E:y} reduces to the condition
$[ 4k(y+\eta_0)y^2 - 2y ]\sinh^2y + y^2\sinh2y = 0$
As described by \eqref{E:curvey}, $y=0$ is an extraneous root since $\eta_0>0$, and
\eqref{E:y} is satisfied if and only if
	\begin{equation}
	\frac{y}{\beta}=\frac{bQ^2}{2}\left(-\coth y +\frac{1}{y}\right)
				+Q \left( b\gamma - \frac{h}{2} \right) \equiv g(y).
	\label{E:yNew}
	\end{equation}
As was mentioned in \lemRef{L:converge}, any solution $y_0$ must be negative since 
$\eta_0>0$.  The function $g(y)$ on the right side of \eqref{E:yNew} 
decreases from $bQ^2/2 + Q(b\gamma - h/2)$ to $Q(b\gamma -h/2)<0$  as $y$
increases from $-\infty$ to zero.  Therefore $g(y)$ intersects $h(y)=y/\beta$ at
exactly one point $y_0<0$.
This shows that there is a unique solution $y_0<0$ to
\eqref{E:yNew}.  Consequently, the global maximum of $|f|^2$ on $z(s)$
occurs at $(0,y_z(0))$, the $y$-intercept of the curve $z(s)$.
\end{proof} 

Combining the result of \lemRef{L:saddleRegion} with \lemRef{L:converge}, it is seen
that
	\begin{equation}
	F(\beta,a,b,c,\qmin,\qmax) = 
		\lim_{\np\to\infty} \frac{1}{\beta \np} 
			\ln( \mathcal{Z}_{\np,\beta,a,b,c,\qmin,\qmax} )
		= - \gamma(b\gamma-h)
		 +\frac{\ln Q}{\beta}
		 +\frac{1}{\beta}\ln\left[ f(0,y_z(0)) \right],
	\label{E:cournotPartCalc0}
	\end{equation}
and the unique global maximum of $|f|$ at $(0,y_z(0))$ on the curve $z(s)$ excludes
any phase transition for this antiferromagnetic model.

Finally, notice that \lemRef{L:saddleRegion} shows that the only extreme point
of $|f|^2$ on the $y$-axis is at the $y$-intercept of the curve $z(s)$.  This
extreme point must be a saddle point of the holomorphic function $f(z)$, hence
$(0,y_z(0))$ is a global minimum of $v(y)=f(0,y)$.  An explicit formula for the
Cournot free energy can then be written
	\begin{equation}
	\begin{aligned}
	F(\beta,a,b,c,\qmin,\qmax)   
	=&-\frac{1}{4} (\qmax+\qmin)[ b(\qmax+\qmin)- 2(a-c) ]
	  +\frac{\ln(\qmax-\qmin)}{\beta}
	\\
	&+ \frac{1}{\beta}\min_{y\in(-\infty,\infty)}
		-\beta b y^2 
		+ \ln\left(\sinhc\left[ 
				\beta b(\qmax-\qmin)y + 
				\beta\frac{(\qmax-\qmin)}{2}\{b(\qmax+\qmin)-(a-c)\}
				\right]
			\right)
	\end{aligned}
	\label{E:cournotPartCalc} 
	\end{equation}
where $y$ was translated and rescaled (this does not alter the result since $y$ ranges over all real numbers) to get rid of square roots, and $\sinhc(u)=\sinh(u)/u$ with
$\sinhc(0)=1$.

Now for a final note on the expected value of the variable $q_i$.  The minimum value in
\eqref{E:cournotPartCalc} occurs when the derivative of the smooth function within the
minimum is zero, which is at $y_m=y_z(0)$.  Since the minimum is a global minimum,
the second derivative with respect to $y$ is nonzero.  By the implicit function
theorem, the minimum point $y_m$ is locally a smooth function of $h=a-c$.
As a result, the partition function in the form \eqref{E:cournotPartCalc0} is seen
to be a smooth function of $h$.

The function $F_\np(h)$ in \eqref{E:freeEnergy} is convex in $h$ via Holder's
inequality, and as a result 
	\begin{equation}
	\frac{F_\np(h)-F_\np(h-\delta)}{\delta} 
	\leq F'_\np(h)
	\leq \frac{F_\np(h+\delta)-F_\np(h)}{\delta}
	\label{E:mag}
	\end{equation}
where $\delta>0$.  
Since $F(h)$ is a smooth function of $h$, taking the limit of \eqref{E:mag} as $\np\to\infty$
results in $F'(h)=\lim_{\np\to\infty}F'_\np(h)$.
Using \eqref{E:par3}
	\begin{equation}
	F'_{\np}(h) = \gamma - \eta/b - i\brfn{\hat{t}}/(2\beta\sqrt{b}),
	\label{E:mag2}
	\end{equation} 
where $\brfn{\cdot}$ is a (one-dimensional) Fourier transform of the Gibbs measure for
$\np$ agents,
$\brn{f}=(\int f\,\e^{\beta V})/\mathcal{Z}_\np$, which is generated by \eqref{E:partitionCO2}.
The measure $\brfn{\cdot}$ is generated by the form of the partition function in
\eqref{E:par3}.  
Taking the limit of \eqref{E:mag2} as $\np\to\infty$,
	\begin{equation} 
	F'(h)= h/(2b) + \lim_{\np\to\infty}\brfn{2t/(Q\sqrt{b})}/(2\beta\sqrt{b}).
	\end{equation}  
The latter limit can be evaluated by deforming the integral in the complex plane as before
(change $t$ to $z$), and using \lemRef{L:converge}.
The sequence $\brfn{\cdot}$ of measures converges to a `delta function' or evaluation measure 
at $(0,y_m)$.  Such measures are precisely the \emph{characters}, meaning they are homomorphisms
on the `algebra of observables' and as such are extreme points.  This shows that there is
no phase transition in this antiferromagnetic model.
It is worthwhile to elaborate on such technicalities to show that the weak convergence of the
Fourier transformed measures implies that the sequence of measures $\brn{\cdot}$
converges weakly.  With \lemRef{L:gibbsConv}, this implies the Gibbs measures
$\brn{\cdot}$ converge weakly to the infinite-agent measure $\br{\cdot}$.

%
%
%
%
%
%
%
\begin{lemma} \label{L:gibbsConv}
The weak convergence of the sequence $\brfn{\cdot}$ of measures
    \footnote{This is the one-dimensional Fourier transform of the Gibbs 
    		  measures $\brn{\cdot}$ in the variable $s_\np\equiv(q_1+\cdots+q_\np)/\sqrt{n}$
    		  that is generated by the partition function in \eqref{E:par3}}
implies the weak convergence of the sequence of Gibbs measures $\brn{\cdot}$.
\end{lemma}
\begin{proof}
Let $\mathfrak{A}_\np$ be the \emph{algebra of observables for \np\ agents}, which is simply
$L^\infty([\qmin,\qmax]^\np, dq_1\cdot dq_\np)$.
The algebra of observables (functions) for an infinite number of agents, 
$\mathfrak{A}_\infty$, is the inductive limit (von Neumann algebra sense) of the finite-agent
algebras $\mathfrak{A}_\np$.  
As a result, the $\mathfrak{A}_\np$ can be considered subalgebras of $\mathfrak{A}_\infty$
and any weak limit of finite-agent measures is defined on $\mathfrak{A}_\infty$.

Consider two distinct limit points $\br{\cdot},\br{\cdot}'\in\mathfrak{A}_\infty^*$
which are limits of subsequences $\brn[1]{\cdot}$ and $\brn[2]{\cdot}$, respectively.
There is a function $g\in\mathfrak{A}_\infty$ on which $\br{\cdot}$ and $\br{\cdot}'$ differ.
By approximating $g$ closely enough with a function in some $\mathfrak{A}_{n_0}$ we may assume
$g\in\mathfrak{A}_{n_0}$.

The integral $\brn[i]{g}$ over the $q_j$, $1\leq j\leq n_i$ can be manipulated into an integral
over the variable $\hat{t}$ as was done with $\eqref{E:par3}$, upon which
it becomes an integral $\brfn[i]{G}$, where $G=G(\hat{t})$ is a function independent of $\np_i$.
This shows $\br{g}=\lim_{\np_1}\brfn[1]{G}$ and $\br{g}'=\lim_{\np_2}\brfn[2]{G}$, 
which implies the assertion. 
\end{proof}

\begin{remark}
With $\vec{q}=\vq$, $g=g(\vec{q})$, note that \eqref{E:approx} inserts the Fourier transform 
$\mathfrak{F}$ (of the Gibbs measure) 
into the $\brn{g}$ integral, and \eqref{E:par2} shows what the dual of the 
Fourier transform, $\mathfrak{F}^*$ does to $g$.
	\footnote{In particular, $\mathfrak{F}^*[g](\sqrt{b}t)= \int \exp(i\sqrt{b}t s_\np) 		    E[g(s_\np,\mathbf{s}_\np^\perp)\chi_Q(s,\mathbf{s}_\np^\perp)|s_\np]d s_\np$,
		    where $\mathbf{s}_\np^\perp$ are integration variables orthogonal
		    to $s_\np$, $\chi_Q$ is the indicator function on
		    $\vec{q}\in[-Q/2,Q/2]^\np$, and $E[\cdot|s_\np]$ is the conditional expectation
		    over the subalgebra of functions in the variables $\mathbf{s}_\np^\perp$.}
The Fourier transform of the Gibbs measures need only converge on the subalgebra of
functions generated by the $\mathfrak{F}^*[\mathfrak{A}_\np]$ in order for the sequence of
Gibbs measures to converge.
\end{remark}

The expected value $\br{q_i}$ is then
	\begin{equation}
	F'(h) = \frac{h}{2b} + \frac{y_m(\beta,b,h)}{\beta Qb}.
	\label{E:magnetization}
	\end{equation}

From \eqref{E:yNew}, we can see that
	\begin{equation}
	\lim_{\beta\to 0} F'(h) = \frac{h}{2b} + \left(\gamma - \frac{h}{2b}\right) 
					= \gamma,
	\label{E:magInfTemp}
	\end{equation}
which simply states that for completely irrational behavior (i.e.,`high temperature'),
the agents will act randomly, and the Gibbs-Cournot measure will be uniform.  As such,
the expected value $\br{q_i}$ over the interval $[\qmin,\qmax]$ is the average
$\gamma=(\qmax+\qmin)/2$.

In the completely rational limit (i.e., `zero temperature'), if $Q/2 + \gamma -h/(2b)>0$
and $\eta\leq0$ (equivalently, if $\gamma\leq h/(2b)<\qmax$),
then the curve $g(y)$ which equals the right side of \eqref{E:yNew} has a positive limit
as $y\to-\infty$.  In this case, the line $h(y)=y/\beta$ intersects $g(y)$ at a point
with $y$-value $y_m$ which has a finite limit as $\beta\to\infty$.  Thus
$y_m/\beta\to 0$ as $\beta\to\infty$ and
	\begin{equation}
	\lim_{\beta\to\infty} F'(h) = \frac{h}{2b} 
					\qquad\qquad\qquad\text{for\ }\gamma\leq h/(2b)<\qmax.
	\label{E:magInfTempHsmall}
	\end{equation}
In the case when $h/(2b)\geq\qmax$, the curve $g(y)$ is always
negative-valued.  As a result, $y_m\to-\infty$ as $\beta\to\infty$.  In this case,
$g(y)$ can be used to evaluate $y_m/\beta$ and 
	\begin{equation}
	\lim_{\beta\to\infty} F'(h) = \frac{h}{2b} + 
			\left( \frac{Q}{2} + \gamma -\frac{h}{2b} \right)
	       = \qmax \qquad\qquad\text{for\ } h/(2b)\geq\qmax.
	\label{E:magInfTempHlarge}
	\end{equation}
Since the integral in \eqref{E:par3} is even in $\eta=b\gamma-h/2$,
$\lim_{\beta\to\infty} F'(h)=\qmin$ when $h/(2b)\leq\qmin$.
If $b\gamma-h/2>0$, then $g(y)$ in \eqref{E:yNew} is shifted up and $g(0)>0$.  With this,
$g(y)$ can only intersect the line $y/\beta$ at positive $y_m$.  It is then seen that
$y_m>0$ when $h/(2b)<\gamma$ and $y_m\leq 0$ when $h/(2b)\geq\gamma$.

In summary, we see that the expected value $\br{q_i}$ in the irrational case is the average
value $(\qmax+\qmin)/2$.  In the completely rational case, we recover the classical
Nash equilibrium value $\br{q_i}=h/(2b)$ when $\qmin\leq h/(2b)\leq\qmax$. When $h/(2b)<\qmin$ 
($>\qmax$) then $\br{q_i}$ will be $\qmin$ ($\qmax$).
For finite $\beta$, the equilibrium $\br{q_i}$ will be smaller (larger) than $h/(2b)$ if
$h/(2b)$ itself is larger (smaller) than the average $\gamma=(\qmax+\qmin)/2$.
Furthermore the graphical analysis below \eqref{E:yNew} shows that $|y_m|/\beta$ is a
decreasing function of $\beta$, and $y_m<0(>0)$ when $h/(2b)>\gamma(<\gamma)$.
As a result, the agents deviate farther from the Nash equilibrium $h/(2b)$
and closer to the average $\gamma$ as $\beta$ decreases 
(i.e., as deviations from rationality increase).  

The \emph{volatility} (i.e., susceptibility)
	\begin{equation}
	\vty=\frac{1}{\beta}F''(h)
	=\lim_{n\to\infty} \frac{1}{\np} \sum_{i,j=1}^{\np}\br{q_iq_j}-\br{q_i}\br{q_j}
	\end{equation} 
can be evaluated in a similar manner.  Let
	\begin{equation}
	G(h,y) = \frac{bQ^2}{2}\left(-\coth y +\frac{1}{y}\right)
				+Q \left( b\gamma - \frac{h}{2} \right)
				-\frac{y}{\beta},
	\end{equation}
which equals zero at points $(h, y_m(h))$.
Then
	\begin{equation}
	\frac{1}{\beta}\frac{dy_m}{dh} =
	\frac{Q}{\beta bQ^2 (\sinh^{-2}y_m - y_m^{-2}) - 2}.
	\label{E:dydh}
	\end{equation}
In the irrational limit, $\lim_{\beta\to 0}y_m=0$ and
$\lim_{\beta\to 0} =(\sinh^{-2}y_m - y_m^{-2}) = -1/3$. 
Differentiating \eqref{E:magnetization} results in
	\begin{equation}
	\vty =
		\frac{Q^2(\sinh^{-2}y_m - y_m^{-2})/2}{
		      \beta bQ^2(\sinh^{-2}y_m - y_m^{-2}) - 2 },
	\label{E:vol}
	\end{equation}
and as a result
	\begin{equation}
	\lim_{\beta\to 0} \vty = \frac{Q^2}{12}.
	\label{E:volInfTemp}
	\end{equation}

In the completely rational limit, either $y_m$ converges to a finite value 
(when $(\qmin+\qmax)/2<h/(2b)<\qmax$) or goes to $-\infty$ (when $h/(2b)\geq\qmax$). 
If $y_m$ converges to a finite limit, then the denominator of \eqref{E:vol} goes to
$-\infty$ and $\lim_{\beta\to\infty}\vty=0$.  If $y_m$ goes to $-\infty$,
then $\vty\leq(Q^2/4)|\sinh^{-2}y_m - y_m^{-2}|$ and 
$\lim_{\beta\to\infty}\vty=0$.  Using that \eqref{E:par3} in even in $\eta$ as before,
	\begin{equation}
	\lim_{\beta\to\infty} \vty = 0.
	\label{E:volZeroTemp}
	\end{equation}

In summary, when agents behave randomly in the irrational limit, the volatility is simply the variance of a uniform random variable.  Each agent becomes decorrolated from the other agents and the volatility is the limiting-average of the volatility of each agent.  When agents
are completely rational they all output at the Nash equilibrium values and do not deviate
from that, which is reflected by zero volatility.

%
%
%
%
%
%

\end{document}